\documentclass[prl,a4paper,10pt,twocolumn,showpacs,floatfix,nofootinbib,superscriptaddress,amsmath,amsfonts,amssymb,preprintnumbers,longbibliography]{revtex4-1} %superbib,citeautoscript
\setcitestyle{super}

\usepackage[T1]{fontenc}
\usepackage[utf8]{inputenc} %utf8 or latin1
\usepackage{dcolumn,graphicx,color,booktabs,microtype,afterpage}
\graphicspath{{./}{figure/}}

\usepackage{bm}
\usepackage[charter,greekuppercase=italicized]{mathdesign}
\usepackage{xfrac} % for sfrac
\usepackage{siunitx} % align number at decimal
\usepackage[colorlinks,plainpages=false,linkcolor=blue,urlcolor=blue,citecolor=blue,pdfpagemode=UseNone,pdfstartview=FitBH]{hyperref}
\usepackage{paralist}
\usepackage{booktabs}

\makeatletter
\renewcommand\section{\@startsection {section}{1}{\z@}%
   {-2.5ex \@plus -1ex \@minus -.2ex}%
   {0.4ex \@plus.2ex}%
  {\normalfont\large\bfseries}}
\makeatother
\makeatletter
\renewcommand\subsection{%
  \@startsection{subsubsection}{3}{\z@}{3.25ex \@plus1ex \@minus.2ex}%
  {-1em}{\normalfont\normalsize\bfseries}}
\makeatother

\renewcommand{\figurename}{Figure}
\renewcommand{\tablename}{Table}
\makeatletter\renewcommand{\fnum@figure}[1]{\textbf{\figurename~\thefigure~|\ }}\makeatother
\makeatletter\renewcommand{\fnum@table}[1]{\tablename~\thetable.}\makeatother

\newcount\hh \newcount\mm
\hh=\time \divide\hh by 60
\mm=\hh \multiply\mm by 60 \mm=-\mm
\advance\mm by \time
\def\now{\number\hh:\ifnum\mm<10{}0\fi\number\mm}

\definecolor{dgreen}{RGB}{0,127,0}

\newcommand{\chemElement}[1]{\text{#1}}
\newcommand{\BCPO}{$\chemElement{BiCu}_2\chemElement{PO}_6$}
\newcommand{\nwu}[2]{#1\ \mathrm{#2}} % number with unit
\renewcommand{\vec}[1]{\bm{#1}} % vector
\newcommand{\dya}[1]{\bm{#1}} % dyadic

\newcommand{\mphroc}[1]{#1}
\newcommand{\mpcc}[1]{#1}
\newcommand{\mpc}[1]{#1}
\newcommand{\tsc}[1]{#1}
\newcommand{\aprc}[1]{#1}
\newcommand{\fcc}[1]{#1}

\newcommand{\refpart}[2]{\hyperref[#1]{\ref*{#1}(#2)}}

\newcommand{\CasolaPRLcite}[1]{\cite{#1, Casola2013, *[{see also additional information in }]Casola2013arXiv}}

\newcommand{\PlumbNatPhyscite}[1]{\cite{#1, Plumb2016, *[{see also additional information in }]Plumb2014arXiv}}

\begin{document}

\makeatletter\renewcommand{\ps@plain}{%
\def\@evenhead{\hfill\itshape\rightmark}%
\def\@oddhead{\itshape\leftmark\hfill}%
\renewcommand{\@evenfoot}{\hfill\small{--~\thepage~--}\hfill}%
\renewcommand{\@oddfoot}{\hfill\small{--~\thepage~--}\hfill}%
}\makeatother\pagestyle{plain}

%\preprint{\textit{Preprint: \today, \now. For internal use only, do not distribute.}}%\linenumbers

\title{Two \mphroc{Coupled} Chains are Simpler than One: Field-induced Chirality in a Frustrated Quantum Spin Ladder}

\author{M.\,Pikulski}
\affiliation{Laboratory for Solid State Physics, ETH Z\"urich, CH-8093 Z\"urich, Switzerland}

\author{T.\,Shiroka}\email[Corresponding author: \vspace{8pt}]{tshiroka@phys.ethz.ch}
\affiliation{Laboratory for Solid State Physics, ETH Z\"urich, CH-8093 Z\"urich, Switzerland}
\affiliation{Paul Scherrer Institut, CH-5232 Villigen PSI, Switzerland}

\author{F.\,Casola}
\altaffiliation[Current affiliation: ]{Harvard-Smithsonian Center for Astrophysics, Harvard University, Cambridge, MA, 02138, USA \vspace{8pt}}
\affiliation{Laboratory for Solid State Physics, ETH Z\"urich, CH-8093 Z\"urich, Switzerland}

\author{A.\,P.\,Reyes}
\affiliation{National High Magnetic Field Laboratory, Florida State University, Tallahassee, FL 32310, USA}

\author{P.\,L.\,Kuhns}
\affiliation{National High Magnetic Field Laboratory, Florida State University, Tallahassee, FL 32310, USA}

\author{S.\,Wang}
\affiliation{Paul Scherrer Institut, CH-5232 Villigen PSI, Switzerland}
\affiliation{Laboratory for Quantum Magnetism, Ecole Polytechnique F\'ed\'erale de Lausanne, CH-1015 Lausanne, Switzerland}

\author{H.-R.\,Ott}
\affiliation{Laboratory for Solid State Physics, ETH Z\"urich, CH-8093 Z\"urich, Switzerland}
\affiliation{Paul Scherrer Institut, CH-5232 Villigen PSI, Switzerland}

\author{J.\,Mesot}
\affiliation{Laboratory for Solid State Physics, ETH Z\"urich, CH-8093 Z\"urich, Switzerland}
\affiliation{Paul Scherrer Institut, CH-5232 Villigen PSI, Switzerland}

\begin{abstract} %\parfillskip=0pt\relax%\linenumbers
\noindent 

\mpcc{Although the frustrated spin chain (zigzag chain) is a Drosophila of frustrated magnetism, the understanding of a pair of coupled zigzag chains (frustrated spin ladder) in a magnetic field is incomplete. We address this problem through nuclear magnetic resonance (NMR) experiments on \BCPO{} in magnetic fields up to $\nwu{45}{T}$, revealing a field-induced spiral magnetic structure. Conjointly, we present advanced numerical calculations showing that even moderate rung coupling dramatically simplifies the phase diagram below half-saturation magnetization by stabilizing a field-induced chiral phase. Surprisingly for a one-dimensional model, this phase and its response to Dzyaloshinskii-Moriya (DM) interactions adhere to classical expectations. While explaining the behavior at the highest accessible magnetic fields, our results imply a different origin for the solitonic phases occurring at lower fields in \BCPO. An exciting possibility is that the known, DM-mediated coupling between chirality and crystal lattice gives rise to a new kind of spin-Peierls instability.}
\end{abstract}


\maketitle\enlargethispage{3pt}

\begin{figure}[ht]
\centering
\includegraphics[width=\columnwidth]{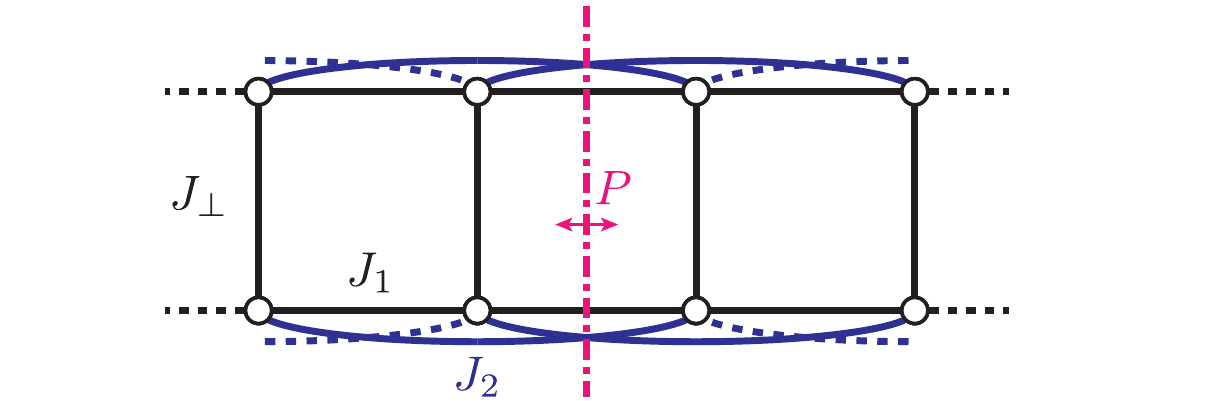}
\caption{\label{fig:frustratedLadder}\textbf{Frustrated spin ladder.} 
A finite segment of the infinitely-extended frustrated spin-ladder model. Vertices correspond to spin-$\sfrac{1}{2}$ magnetic moments, while edges represent the relevant exchange interactions $J_1$, $J_2$, and $J_\perp$. The system exhibits a reflection symmetry denoted \mpc{as} $P$.}
\end{figure}

\noindent{}
\mpcc{Despite previous \mphroc{studies},\CasolaPRLcite{Tsirlin2010,Shyiko2013,Sugimito2014jpsconf,Sugimoto2015}} \mpcc{\mpc{the effects of external magnetic fields on materials} in which \mpc{antiferromagnetic (Heisenberg)} exchange interactions between spin-$\sfrac{1}{2}$ moments form \mpc{quasi one-dimensional} frustrated spin ladders} (Fig.~\ref{fig:frustratedLadder}) \mpcc{are not fully understood}.
\mpcc{The} frustrated ladder model\mpc{\cite{Vekua2006}} encompasses the \mpc{unfrustrated} spin ladder\cite{Dagotto1996} and the spin chain with frustrating next-nearest neighbor (NNN) interactions (zigzag chain)\mpcc{,}\cite{Majumdar1967_1,*Majumdar1967_2} as limiting cases for $J_2 = 0$ and $J_{\perp} = 0$, respectively.\mpc{\cite{Lavarelo2011}}
\mpcc{Given a} \mpc{sufficiently-strong} frustration \mpcc{ratio $J_2/J_1$ in the latter case}, both aforementioned systems adopt \tsc{a} spin-singlet ground \tsc{state} with a spin gap and short-range spin correlations only\mpc{\cite{Shastry1981,Bursill1995,Dagotto1996}}\mpcc{---}\tsc{two} hallmarks of quantum spin liquids (QSLs).\cite{Balents2010}

Closing the gap by applying a magnetic field typically induces magnetic order in such systems.\mpc{\cite{Wessel2000}}
For example, a field-induced Bose-Einstein condensation (BEC) of mobile spin-triplet excitations (triplons) gives rise to \mpc{antiferromagnetic (AFM)} order in the \mpc{unfrustrated} spin ladder with dominant \mpc{rung exchange} $J_\perp$.\cite{Giamarchi2008}
On the other hand, the frustration of the zigzag chain entails dimerization\cite{Nomura1994} and
incommensurate, spiral-like spin correlations,\cite{Chitra1995,Nersesyan1998} resulting in more complicated field-induced phases.\cite{Hikihara2010}
In particular, the zigzag chain supports field-induced chiral order,\cite{Okunishi2008,Kolezhuk2005,McCulloch2008,Hikihara2010} which \mpc{can be described as a condensation of magnetic excitations with incommensurate wavevector.\cite{Ueda2009}}
\tsc{This order} is characterized by translationally-invariant expectation values of the longitudinal component of the chirality operator $\vec{\kappa}_{ij} = \vec{S}_i \times \vec{S}_j$ (cf.~Eq.~7 in \nocite{Villain1977}Ref.~\citenum{Villain1977}) quantifying the ``twist'' of the magnetic structure along a bond $(i,j)$ between two magnetic sites with spin operators $\vec{S}_i$ and $\vec{S}_j$.\mpcc{\cite{Kolezhuk2005}}
Note that chiral order only breaks\cite{Nersesyan1998} the discrete reflection symmetry $P$ (Fig.~\ref{fig:frustratedLadder}) and can therefore occur even in the absence of ordered moments.\cite{Villain1978}

Intuitively, the ordered state corresponds to a spiral structure with fixed handedness (left- or right-handed), but undefined propagation phase.
\mpc{This is plausible, given that such spiral structures form the ground states of the classical zigzag chain ($S \to \infty$) and are, indeed, expected to arise if residual interladder couplings permit conventional long-range magnetic order.\cite{Hikihara2010}}
\mpc{Such spiral order corresponds to a} secondary breaking of the $U(1)$ symmetry emerging\cite{Ueda2009} \mpc{from} the combination of lattice-translation invariance \mpcc{and} \tsc{incommensurate} spin correlations.\footnote{Note that this quasi-continuous $U(1)$-symmetry persists even if the continuous $SU(2)$ symmetry of the magnetic moments themselves is fully lifted, \mpcc{e.\,g.\,, by an} external magnetic field and DM interactions.}
\mpc{Conceptually, this is} similar to the aforementioned appearance of long-range AFM order following a field-induced BEC of triplons in the \mpc{unfrustrated} spin ladder.

\begin{figure}[h]
  \includegraphics[width=\columnwidth]{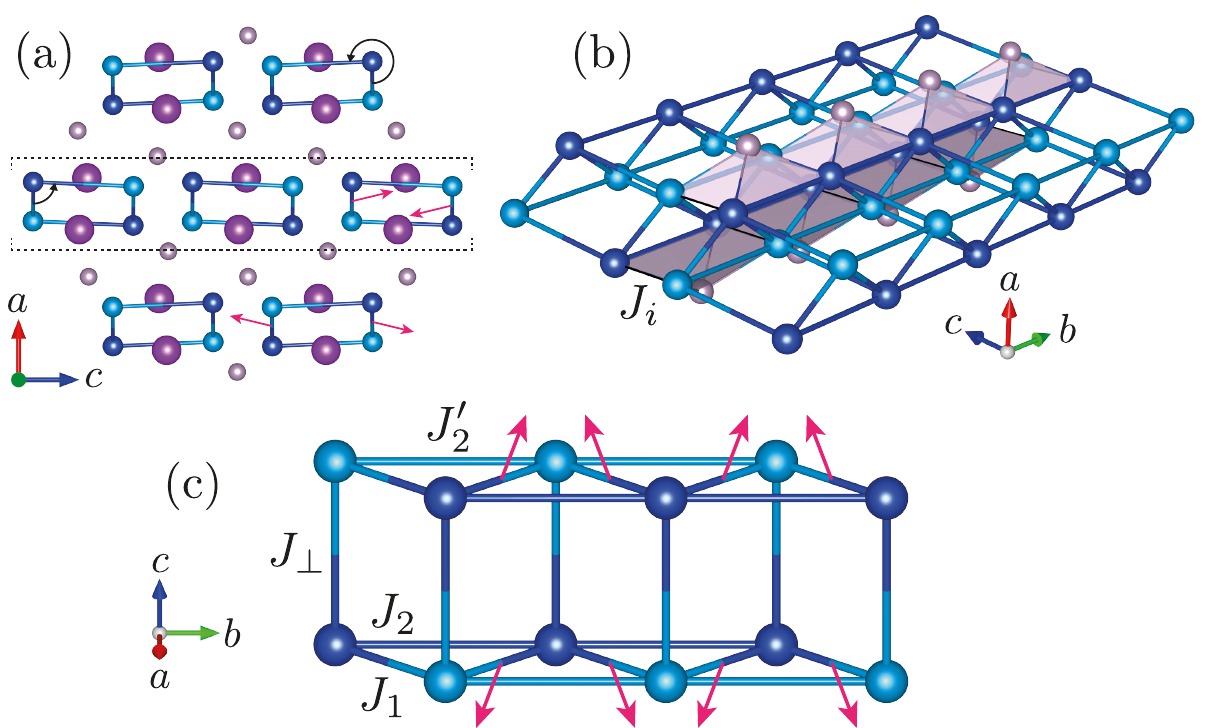}
  \caption{\textbf{Crystal structure of \BCPO{}.} The positions\cite{Abraham1994} of $\chemElement{Cu}$ (light and dark blue), $\chemElement{P}$ (gray) and $\chemElement{Bi}$ (purple) are shown; $\chemElement{O}$ sites have been omitted for clarity. Exchange interactions\cite{Tsirlin2010} ($J_1$, $J_2^\prime$, $J_2$, $J_\perp$, and $J_i$) are depicted by blue (intraladder couplings) and black (interladder couplings) lines. (a) View along $b$, illustrating the two ladder orientations. Dashed lines mark one ladder layer. (b) Two ladder units. Hyperfine couplings\cite{Alexander2010} between $^{31}\chemElement{P}$ nuclei and $\chemElement{Cu}$ sites are represented by gray pyramids. (c) Single ladder. Magenta arrows in (a) and (c) illustrate the staggering of the transverse and parallel components, $D_1^{ac}$ and $D_1^b$, of the DM vectors\mpc{\PlumbNatPhyscite{Hwang2016}} on the nearest-neighbor bonds [projected into $bc$ plane in (c)]. Illustrations created using VESTA.\cite{Momma2011}}\label{fig:structure}
\end{figure}

The zero-field ground-state phase diagram of the frustrated spin ladder (Fig.~\ref{fig:frustratedLadder}) essentially interpolates between the \mpc{unfrustrated} spin ladder and the zigzag chain.\mpc{\cite{Lavarelo2011}}
Specifically, the dimerized character of the zigzag chain is retained for small rung couplings, whereas stronger rung couplings yield a uniform ground state with dominant resonating-valence-bond (RVB) contributions from the rung bonds.\cite{Vekua2006,Lavarelo2011}
As in the zigzag chain, sufficient frustration induces incommensurate spin correlations\cite{Lavarelo2011} \mpc{and chiral order \tsc{thus} becomes possible.\CasolaPRLcite{Shyiko2013}}

\mphroc{Our work aims at understanding} the intriguing variety of field-induced phases \mpc{reported} \mpcc{previously}\CasolaPRLcite{Kohama2012,Kohama2014} for \BCPO{}\mpc{\cite{Abraham1994}}, \mphroc{which is believed to be described by} \mpcc{the frustrated spin-ladder model depicted in \mpcc{Fig.~\ref{fig:structure}}.\PlumbNatPhyscite{Koteswararao2007,Mentre2009,Tsirlin2010,Plumb2013,Hwang2016,Splinter2016}}
\mphroc{We address this question through both }\tsc{\mphroc{new} \mpcc{calculations} and new high-field} \mpcc{experiments}. \mphroc{Moreover, to allow a meaningful comparison with the real compound, our considerations} account for the presence of two inequivalent magnetic sites with corresponding next-nearest neighbor (NNN) couplings ($J_2^\prime$ and $J_2$),\cite{Tsirlin2010} as well as various symmetry-allowed Dzyaloshinskii-Moriya\cite{Dzyaloshinsky1958,Moriya1960_1} (DM) interactions.\PlumbNatPhyscite{Hwang2016}
Previous numerical calculations for \BCPO{} revealed the appearance of field-induced chirality \mpc{for a particular choice of model parameters}.\CasolaPRLcite{}
However, the dependence of the field-induced chiral phase on the exchange couplings and \tsc{on} \mpcc{the} DM interactions was not considered in detail.
\mpcc{The} calculations \mpcc{further} seemed to indicate the presence of another magnetic phase at low system magnetizations. \mpc{On the other hand, similar \mpcc{features} had been attributed to convergence problems in previous work on the zigzag chain.\cite{Sudan2009}}
In the following, we report the results of \tsc{\emph{comprehensive numerical calculations}} for the frustrated ladder model [Fig.~\refpart{fig:structure}{c}], which clarify that \begin{inparaenum}[(i)] \item a field-induced chiral phase generally appears for sufficiently strong frustration and rung coupling, and \item no additional field-induced phases occur at lower magnetizations\end{inparaenum}.
\mpc{Overall, the field-induced phases \mpcc{occurring} below half\mphroc{-}saturation magnetization are much more similar to the classical ground state than \mpcc{for an} isolated zigzag chain.}

\fcc{To \mpcc{connect} our calculations with the experimentally-tangible reality,} we present $^{31}\chemElement{P}$ nuclear magnetic resonance (NMR) data collected on \BCPO{} in high magnetic fields $\vec{H} \parallel b$. For this orientation, magnetic phase transitions were observed at critical fields $\mu_0\, H_{c1} \simeq \nwu{20}{T}$ and $\mu_0\, H_{c2} \simeq \nwu{34}{T}$ (for $T \sim \nwu{0}{K}$).\cite{Kohama2012}
The state above $H_{c1}$ was interpreted as a soliton lattice, and an instability towards chiral order at even higher fields was proposed.\CasolaPRLcite{}
\tsc{Our new data} \mpcc{are consistent with} the latter prediction and \mpcc{we} \tsc{offer new insights} \mpcc{by discussing} the results in view of our \tsc{new} \mpcc{calculations,} as well as other experiments, including measurements of the electric polarization.\cite{Jeon2014}

\begin{figure}[ht]
  \includegraphics[width=\columnwidth]{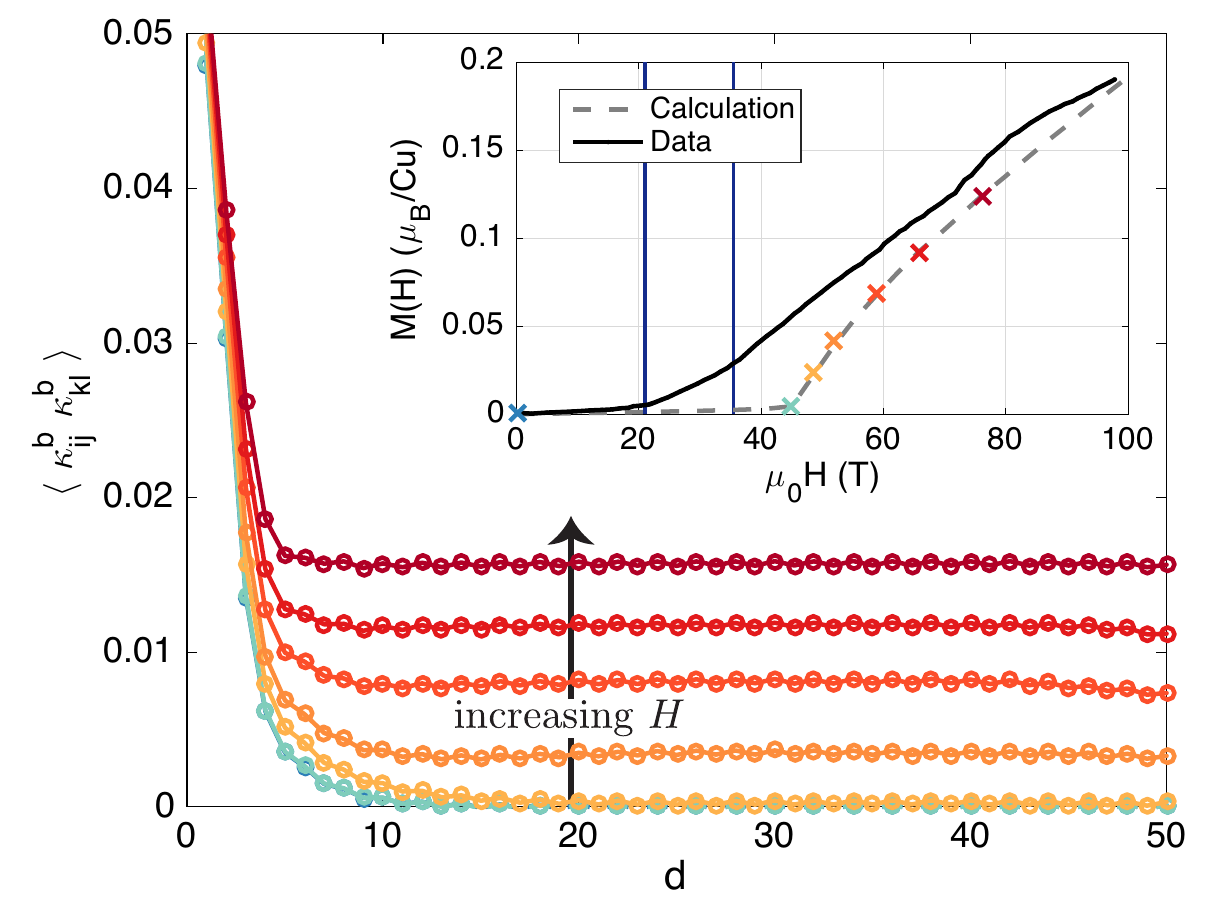}
  \caption{\textbf{Chiral correlations} Correlations of longitudinal chirality $\kappa_{ij}^b$ on two nearest-neighbor bonds $\langle ij \rangle$ and $\langle kl \rangle$ as \mpc{a} function of distance $d$ along the ladder leg, calculated for different magnetic fields (cf.~inset). Inset: Calculated magnetization, along with data reported in Ref.~\citenum{Kohama2014}. Crosses indicate the fields at which the correlation functions shown in the main panel were evaluated and vertical blue lines correspond to \tsc{the experimental} $H_{c1}$ and $H_{c2}$ \tsc{values}.\cite{Kohama2012}}\label{fig:Kzz}
\end{figure}

\section{Results}\label{sec:results}

\subsection{Numerical calculations for the frustrated-ladder model.}\label{sec:calc}
We model the system \tsc{shown in} \mpc{Fig.~\refpart{fig:structure}{c}} using the spin Hamiltonian\PlumbNatPhyscite{}
\begin{equation}\label{eq:Hamiltonian}
  H = \sum_{\langle ij \rangle} \left[ J_{ij}\; \vec{S}_i \cdot \vec{S}_j
    + \vec{D}_{ij} \cdot \left( \vec{S}_i \times \vec{S}_j \right) 
    + \vec{S}_i \cdot \dya{\Gamma}_{ij} \cdot \vec{S}_j \right] \quad,
\end{equation}
where $\langle ij \rangle$ iterates over pairs of interacting sites $i$ and $j$, and $\vec{D}_{ij}$ denotes the DM vectors.
The symmetric tensor $\dya{\Gamma}_{ij}(\vec{D}_{ij})$ arises for anisotropic superexchange interactions.\cite{Kaplan1983,Shektman1992}
We assume that $\vec{H} \parallel b$ \mpc{in the experiment, which replaces the two inequivalent g-tensors by scalars. Further details regarding the modeling are described in Methods.}

Calculated correlation functions of the longitudinal chirality are shown in Fig.~\ref{fig:Kzz}. The correlations are short ranged until the applied magnetic field is large enough to suppress the spin gap (kink in the calculated magnetization) and become long ranged immediately thereafter. In accordance with an incipient order with spiraling transverse magnetic moments, the asymptotic behavior of the transverse spin correlations concomitantly switches from \mpc{short-ranged} to \mpc{slowly-decaying}.

\begin{figure}[h]
  \includegraphics[width=\columnwidth]{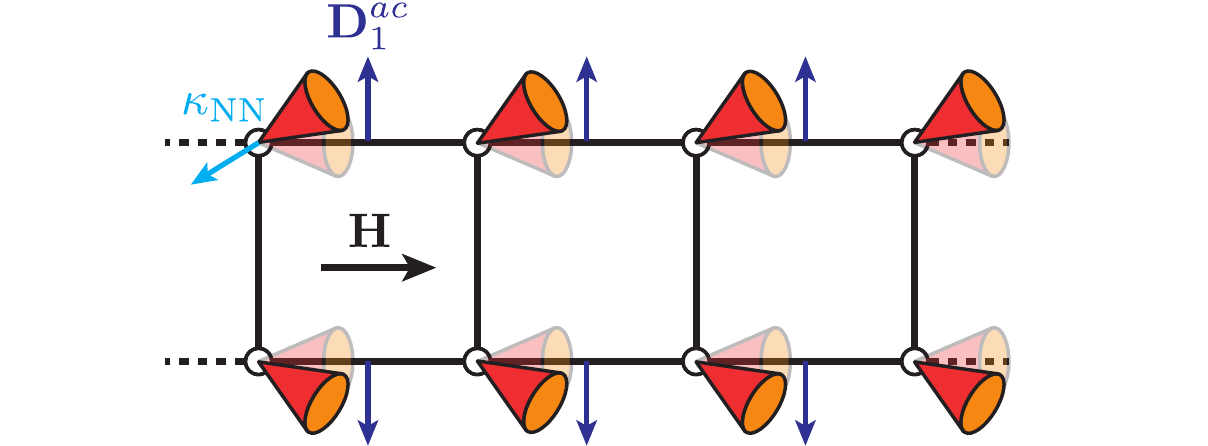}
  \caption{\textbf{Spiral structure with DM-induced distortions.} Schematic illustration of tilt distortion expected due to the transverse DM interaction $\vec{D}_1^{ac}$ [Fig.~\refpart{fig:structure}{c}]. The cones represent canted spiraling magnetic moments with incommensurate propagation along the leg direction and $\vec{\kappa}_\text{NN}$ denotes the vector chirality on the nearest-neighbor leg bonds. Opaque (transparent) colors are used to depict the situation with (without) DM interactions.}\label{fig:tilt}
\end{figure}

The \mpc{above behavior is} robust against moderate DM interactions (see Methods).
Since the DM interactions can be written as $\vec{D}_{ij} \cdot \vec{\kappa}_{ij}$, it is natural to expect DM-induced twist and tilt distortions for classical spiral structures \mpc{(see, e.\,g.\,, Ref.~\citenum{Veillette2005})}. The latter case is illustrated in Fig.~\ref{fig:tilt}.
Indeed, we found the correlation functions in the field-induced chiral phase of the ideal one-dimensional quantum system to be altered accordingly in the presence of DM interactions.

\begin{figure}[t]
  \centering
  \includegraphics[width=0.95\columnwidth]{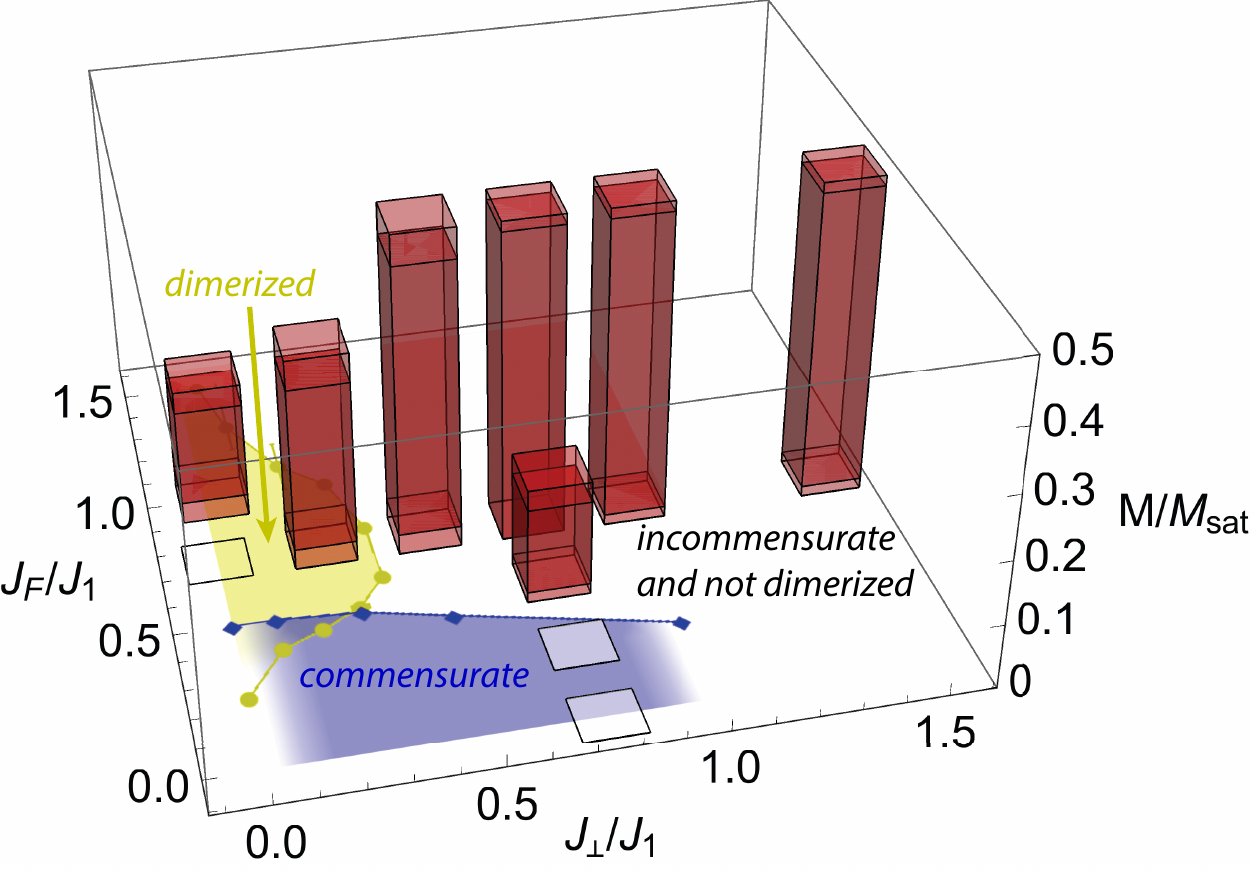}
  \caption{\textbf{Influence of exchange couplings.} Magnetization region occupied by the chiral phase \mpc{(red bars)}, as \mpc{a} function of exchange couplings [$J_F = (J_2 + J_2^\prime)/2$, $J_2 / J_2^\prime = 2$ fixed]. The magnetization $M$ is normalized to the saturation magnetization $M_\text{sat}$. Transparency is used to indicate uncertainties. \mpc{The zero-field phase boundaries towards the dimerized (light green) and commensurate phases (blue) of the frustrated ladder with $J_2=J_2^\prime$,\cite{Lavarelo2011} redrawn using data points provided by the authors of Ref.~\citenum{Lavarelo2011}, are included for comparison.}}\label{fig:phaseDiagramChirality}
\end{figure}

The influence of the exchange interactions, in the absence of DM interactions, is illustrated in Fig.~\ref{fig:phaseDiagramChirality}. 
\mpc{Note that we assumed two inequivalent NNN bonds ($J_2^\prime \neq J_2$) for consistency with our experimental work on \BCPO{}, which deviates from $J_2^\prime = J_2$ used in Ref.~\citenum{Lavarelo2011}. However, besides allowing for \mpcc{different} ordered moments on the two magnetic sites, we find no effects of this assumption on the zero-field phase diagram or the field-induced chiral phase mapped in Fig.~\ref{fig:phaseDiagramChirality} (see Methods).}

\begin{figure*}[t]
  \includegraphics[width=2\columnwidth]{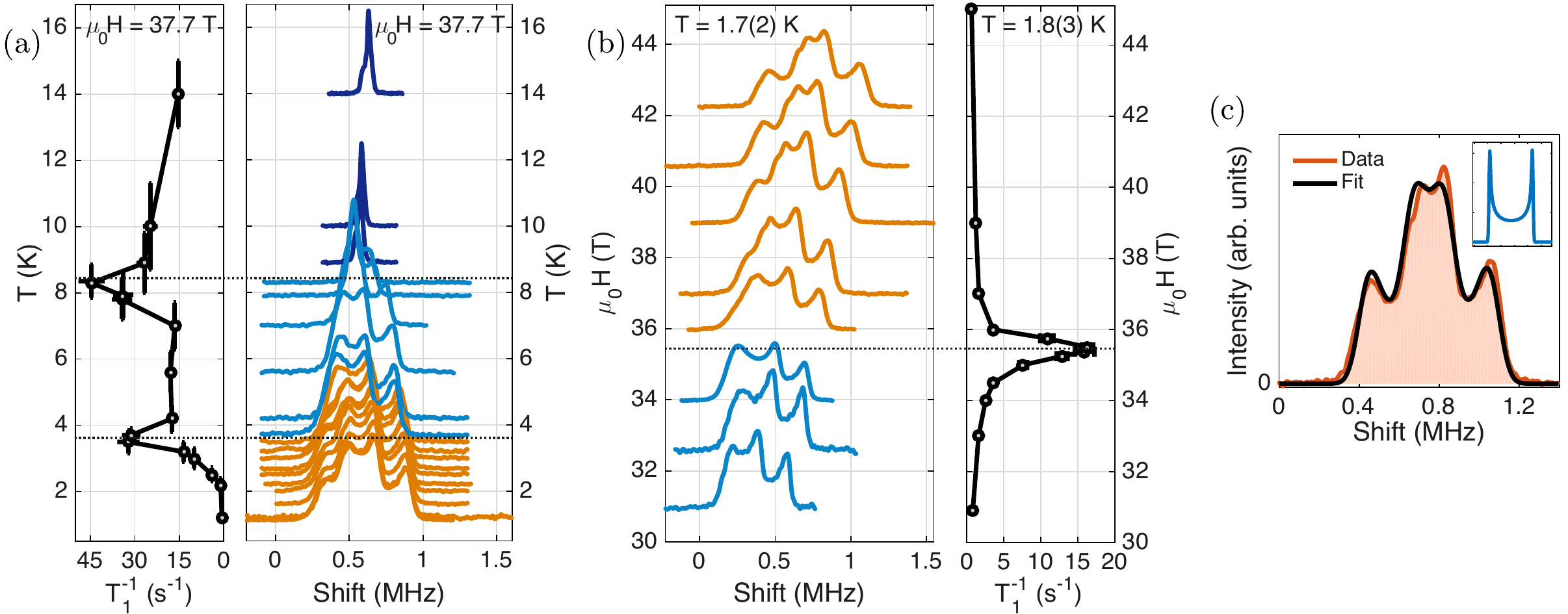}
  \caption{\textbf{High-field NMR data.} (a-b): $^{31}\chemElement{P}$-NMR spectra and relaxation rates $T_1^{-1}$ of \BCPO{}, as function of (a) temperature $T$ and (b) magnetic field $\mu_0\, H$ ($\vec{H} \parallel b$). The baseline ordinates of the spectra encode the corresponding temperatures and magnetic fields, respectively. Spectra are normalized to their maximal intensities. Colors distinguish different phases and dotted lines indicate the approximate locations of the peaks in $T_1^{-1}$. (c): $^{31}\chemElement{P}$-NMR spectrum measured at $\mu_0\, H = \nwu{42.2}{T}$ and $T = \nwu{1.7}{K}$, along with \tsc{the} fitted line shape. The inset shows \mpc{a typical} double-horn spectrum.}\label{fig:spectraT1}
\end{figure*}

\subsection{High-field NMR.}\label{sec:nmr}

Figure~\refpart{fig:spectraT1}{a-b} shows the measured \mpc{NMR} spectra and relaxation rates.
\mpc{The effects of decreasing temperature in high fields (a) and of increasing field at low temperature (b) are similar.}
The $^{31}\chemElement{P}$-NMR spectrum, which reflects \mpc{the} distribution of internal magnetic fields at the $\chemElement{P}$ site, first evolves from a single line (dark blue spectra) to a three-peak structure (light blue spectra) reported\CasolaPRLcite{} earlier (see Methods).
At even higher fields, a fourth peak develops (orange spectra). 
These distinct changes are \mpc{accompanied by} peaks in \mpc{the spin-lattice relaxation rate} $T_1^{-1}$, which are indicative of slow magnetic fluctuations.
Together with the smooth evolution of the spectra, this confirms\cite{Kohama2012} the presence of two second-order phase transitions at temperatures and fields consistent with Ref.~\citenum{Kohama2012}.

\section{Discussion}\label{sec:discussion}

\mpc{Based on the behavior of the calculated correlation functions (cf.\ Fig.~\ref{fig:Kzz}), a field-induced chiral---or, accordingly, spiral---phase is the only phase expected to appear at magnetizations which are of experimental relevance for \BCPO{}.
More generally, Fig.~\ref{fig:phaseDiagramChirality} admits the} conjecture that a direct field-induced transition from the spin-singlet ground state to a chiral phase occurs for all sets of exchange couplings giving rise to a non-dimerized ground state with incommensurate spin correlations. The latter aspect is consistent with the fact that the breaking of the \mpcc{reflection} symmetry $P$ requires a ground state with low-symmetry lattice momentum,\cite{Ueda2009} whereas the former condition suggests that dimerization and chirality are competing phenomena.
Comparison with the phase diagram\cite{Hikihara2010} of the \mpc{isolated} zigzag chain \mpc{($J_\perp = 0$)} finally shows that the field-induced chiral phase of the frustrated spin ladder is connected to that of the zigzag \mpc{chain} in the space of exchange couplings.
\mpc{Moreover, the rung coupling is found to simplify the phase diagram below half\mphroc{-}saturation magnetization by suppressing the various competing phases in favor of a single field-induced chiral phase.
Remarkably, due to the similarity between chiral order and the classically-expected spiral structures revisited in the introduction, the frustrated two-leg ladder thus turns out to be ``much more classical'' than an individual zigzag chain. This observation is further corroborated by \mpcc{the fact that} the DM-induced changes to the correlation functions \tsc{are} consistent with those expected for classical spiral structures.}

After these general results for the frustrated ladder system, we discuss the high-field NMR spectra obtained in \BCPO{} in more detail.
Based on the calculations reported in Ref.~\citenum{Casola2013arXiv} and this work, a field-induced incommensurate spiral structure---possibly with DM-induced distortions---is expected to appear in \BCPO{}.
With the usual assumption of linear hyperfine couplings, such magnetic structures quite generally give rise to double-horn spectra like \mpc{the one} depicted in the inset of Fig.~\refpart{fig:spectraT1}{c}.\cite{Blinc1981}
Indeed, after allowing for a Gaussian broadening, the experimental data can be fitted using a symmetric superposition of two double-horn contributions [Fig.~\refpart{fig:spectraT1}{c}].\footnote{\label{note:narrowComponentFeatures}The additional sub-structure of the narrow spectral component is related to the sample alignment and therefore not considered further (see Methods).}

Given the absence of contrary indications, the magnetic unit cell is taken to coincide with the crystallographic one.
There are four translationally-inequivalent $\chemElement{P}$ sites, each of which is expected to yield an individual double-horn contribution to the NMR spectrum when a sinusoidal magnetic order with incommensurate variation along $b$ is adopted.
These sites are related by three space-group reflections.
Hence---although it incidentally agrees with previous simulations\CasolaPRLcite{}---, the observation of two double-horn contributions with an intensity ratio compatible with ${1 : 1}$ (see Methods) indicates a field-induced breaking of one of these reflection symmetries.

\mpc{Note that \BCPO{} exhibits two types of magnetic layers, which become inequivalent in magnetic fields $\vec{H} \parallel b$ [black arrows in Fig.~\refpart{fig:structure}{a}]. 
This provides a natural explanation for the observed NMR line shape. At the level of} the spin-Hamiltonian governing electronic and nuclear magnetic moments, this scenario corresponds to a significantly nonlinear coupling with the external magnetic field. Even a coupling between the magnetic field and the longitudinal chirality of the magnetic moments is conceivable.
However, in this case\mpc{, the experimentally-observed coincidence of the centers of the two double-horn contributions would be accidental, which is why we now turn to the scenario of negligible non-Zeeman terms.}
While the in-plane \mpc{AFM} couplings $J_\perp$ and $J_i$\PlumbNatPhyscite{Choi2013,Plumb2013} [Fig.~\refpart{fig:structure}{a-b}] constrain the magnetic structure within each magnetic layer, the coupling between adjacent magnetic layers is weak.\cite{Plumb2013}
The observed NMR spectra can then be explained if residual interlayer couplings give rise to a stacking of magnetically-ordered layers which is not symmetric along the \mpc{crystallographic $a$-direction}.
\mpc{While the significance of quantitative parameter estimates is limited, we find that plausible solutions reproducing the data shown in Fig.~\refpart{fig:spectraT1}{c} \emph{exist} in general within the aforementioned class of spiral magnetic structures (see Methods).}
\mpc{Furthermore,} since the dipole fields created by adjacent magnetic layers decay rapidly with distance along $a$, a randomly-stacked structure in which the propagation phase of adjacent magnetic layers differs by $\pm \delta$ (with $\delta$ fixed) \mpc{is, in fact,} sufficient to explain the results.
Thus, the observed NMR spectrum [Fig.~\refpart{fig:spectraT1}{c}] is fully consistent with a spiral magnetic structure, as it is expected to emerge due to field-induced chirality in the frustrated-ladder model in the vicinity of interladder couplings.

An electric polarization oriented predominantly along $a$ has been reported to appear at $H_{c2}$ in an order-\mpc{parameter}-like manner.\cite{Jeon2014}
Indeed, magnets exhibiting spiral order often are improper ferroelectrics.\cite{Cheong2007}
In \BCPO{}, magnetic fields $\vec{H} \parallel b$ preserve a two-fold screw-axis symmetry, which the magnetic order must break in order to induce a polarization perpendicular to $b$.
This corroborates the aforementioned reduced-symmetry stacking of magnetic layers.

Considering the NMR spectra in Fig.~\ref{fig:spectraT1}, we notice \mpc{increasing distortions} upon approaching the \mpc{$H_{c2}$ phase boundary from higher fields or lower temperatures, respectively}, which indicates the appearance of defects (solitons or domain walls). This is consistent with results of entropy measurements,\cite{Kohama2014} \mpcc{while} data from electric-polarization experiments\cite{Jeon2014} suggest that these objects are chiral.\cite{Baryakhtar1983,Furukawa2008}
Their detailed structure or location are hard to infer from NMR data \mpc{alone}, since the hyperfine couplings involve sites with very different propagation phases.
Still, the regularity of the distortions implies that defects appear at well-defined locations with respect to the magnetic structure.
Such localization \mpc{effects} can be due to interladder and/or magneto-elastic\cite{Splinter2016} couplings, as in $\chemElement{CuGeO}_3$.\cite{Khomskii1996,Zang1997,Dobry1997,Sorensen1998,Uhrig1999}

While the numerical predictions of a field-induced chiral phase are consistent with \mpc{the experimental data}, previous numerical signs\cite{Casola2013} of solitonic behavior could not be substantiated. We interpret this as a pointer to the importance of \mpcc{interactions beyond the one-dimensional frustrated-ladder model} at magnetic fields $H \lesssim H_{c2}$ (cf.\ preceding paragraph). \mpcc{Since DM interactions couple chirality and crystal structure,\cite{Onoda2007, Sergienko2006_MC} one exciting possibility is that they give rise to a new, DM-based variation of the spin-Peierls instability with associated solitons known\cite{Uhrig1999} from $\mathrm{CuGeO}_3$.}
Also, \mpcc{the same unknown} interactions \mpcc{may explain} the apparent discrepancies between the calculated and measured\cite{Kohama2014} magnetization curves \mpcc{already discussed in Ref.~\citenum{Tsirlin2010} (see also Ref.~\citenum{DissPikulski})}.
Future experimental identification of these interactions \mpc{is} crucial to unraveling the precise nature \tsc{of} defects in \BCPO{}.

To conclude, \mphroc{we performed numerical calculations which enabled us to} clarify a part of the \mpcc{in-field phase diagram} of the frustrated spin ladder model \mphroc{and presented} \mpcc{experimental} evidence that the field-induced phase observed in \BCPO{} for \mpcc{$H > H_{c2}$ ($\vec{H} \parallel b$)} \mpcc{has} a spiral magnetic structure.
Our numerical results show that this order is driven by the field-induced chirality in the individual one-dimensional ladder units, whereas the presence of spiraling ordered moments \tsc{is a mere} secondary effect due to the presence of suitable interladder couplings.
Whilst the frustration within each ladder leg is essential for \mpc{the} field-induced chirality, the rung bonds of the ladder \mpc{support this effect by weakening} the apparently competing dimer order.
Furthermore, our experiments indicate that the behavior of \BCPO{} at magnetic fields below\mpc{, or comparable to,} $H_{c2}$ is governed by defects \mpc{not} captured by the present one-dimensional model.
High-field diffraction experiments could help \mpc{reveal} the nature of \tsc{such} defects, as well as probe \mpc{the} possible \mpcc{magnetic-structure} and lattice distortions in the field-induced spiral phase.

\vspace{5mm}
\section{Methods}

\subsection{Numerical calculations.}
\footnotesize{%
The Hamiltonian (\ref{eq:Hamiltonian}) is studied using exact diagonalization and density-matrix renormalization group\cite{White1992} (DMRG).
\mpc{Following previous work,\CasolaPRLcite{} we} consider individual spin operators $\vec{S}_i$, the dimer operator $\vec{S}_i \cdot \vec{S}_j$, as well as the chirality $\vec{\kappa}_{ij}$, and compute their correlation functions for individual ladders.
The field-induced phases are identified by \mpc{inspecting} the aforementioned correlation functions \mpc{(see, e.\,g.\,, Ref.~\citenum{Hikihara2010})} and the associated structure factors.
The correlation functions are averaged over several reference sites/bonds close to the center of the ladder.\cite{Dolfi2015}
Convergence\mpc{\cite{Dolfi2014}} of the results as function of matrix bond dimension and number of optimization sweeps is checked.
Moreover, the ground-state degeneracy is accounted for.
\mpc{Finite-size effects\cite{Dolfi2015} were checked by performing calculations for several system sizes. The abscissa range shown in Fig.~\ref{fig:Kzz} was restricted to exclude regions strongly affected by the open system boundaries.}

The symmetric anisotropy tensor $\dya{\Gamma}_{ij}$ is given by\cite{Kaplan1983,Shektman1992}
\begin{equation}\label{eq:Gamma}
  \dya{\Gamma}_{ij} = \frac{\vec{D}_{ij} \otimes \vec{D}_{ij}}{2\; J_{ij}} - \frac{D_{ij}^2}{4\;J_{ij}} \quad.
\end{equation}
Two representative parameter sets---without ($A$) and with ($B$) DM interactions, respectively---are primarily considered in this work:
\begin{center}
\begin{tabular}{p{10mm}p{11mm}
         S[table-format=1.2,table-column-width=8mm]
         S[table-format=1.1,table-column-width=8mm]
         S[table-format=1,table-column-width=8mm]
         S[table-format=1.1,table-column-width=10mm]
         S[table-format=1.1,table-column-width=10mm]
         S[table-format=1.1,table-comparator=false,table-space-text-pre = $\sim\;$~]}
  \toprule
  Set & $J_1/k_B$ & {$J_\perp/J_1$} & {$J_2^\prime/J_1$} & {$J_2/J_1$} & {$D_1^{ac}/J_1$} & {$D_1^b/J_1$} \\
  \midrule
  $A$\cite{Tsirlin2010} & $\nwu{140}{K}$ & 0.75 & 0.5 & 1 & 0 & 0 \\
  $B$\PlumbNatPhyscite{Hwang2016} & $\nwu{116}{K}$ & 1 & 1 & 1 & 0.3 & 0.3 \\
  \bottomrule
\end{tabular}
\end{center}
The onset of chiral order and the absence of additional field-induced phases at low magnetization \mpc{values were} checked by simulating systems consisting of up to $L=256$ (128) rungs using \mpc{bond dimensions of} up to $m=2048$ (512) for parameter set $A$ ($B$).
The results shown in Fig.~\ref{fig:Kzz} were obtained using the more recent\PlumbNatPhyscite{Hwang2016} parameter set $B$.
Small differences between chiral correlations at even and odd distances are due to the staggered DM interactions $D_1^b$ [see Fig.~\refpart{fig:structure}{c}].

In order to make use of $S_\text{tot}^z$ conservation, most calculations were performed with a fixed g-factor ($g$=2). \mpc{Note that the assumption of site-independent g-factors $g_1 = g_2$ is implicit to parameter set $B$.\PlumbNatPhyscite{Hwang2016}} \mpc{For parameter set $A$,} inclusion of the site-dependent g-factors fitted for $\vec{H} \parallel b$ ($g_1 = 1.78$, $g_2 = 2.19$) did not affect any of the reported conclusions. The data shown in Fig.~\ref{fig:phaseDiagramChirality} were obtained by varying parameter set $A$ as indicated in the figure. \mpc{Here, the} uncertainties in the phase boundaries are due to practical limitations (except for parameter set $A$, $L=64$ and $m=512$ \mpc{were} used). \mpc{Taking} $\Delta{}J_F = J_2 - J_2^\prime \to 0$ for parameter set $A$ [\mpc{while} keeping $J_F =(J_2^\prime+J_2)/2$ constant] did not affect the field-induced phases below \mpc{half\mphroc{-}saturation} magnetization and the results obtained for $J_\perp = 0$ were consistent with \mpc{those reported in} Ref.~\citenum{Hikihara2010}.
\mpc{Moreover, the zero-field ground states were consistent with those reported\cite{Lavarelo2011} for $\Delta{}J_F = 0$. Hence,} the comparison of our results with previous results\cite{Lavarelo2011} obtained for the case $J_2^\prime=J_2$ is justified.

To study the effect of the DM interactions, we augmented parameter set $A$ by individual symmetry-allowed DM terms ($D_1^b = 0.2\,J_1$, $D_1^{ac} = 0.35\,J_1$, $D_4^b = 0.2\,J_1$, ${D^\prime}_2^a \approx D_2^a = 0.4\,J_1$, and ${D^\prime}_2^c \approx D_2^c = 0.4\,J_1$; cf.\ \mpc{Refs.}~\citenum{DissPikulski},\citenum{Hwang2016}). 
For historical reasons and following standard practice,\cite{Oshikawa1997,Cepas2001,Chernyshev2005,Fouet2006,Hao2011} we set $\dya{\Gamma}_{ij}=0$ in these calculations. In all considered cases, the field-induced chiral phase persisted \mpc{up to magnetizations of at least $\sim 20\%$ of the} saturation magnetization.

Additional results and details are \mpc{documented} in Ref.~\citenum{DissPikulski}.
}

\subsection{High-field NMR: Experimental details.}
\footnotesize{%
The NMR shifts are reported \mpc{relative to} a standard $^{31}\chemElement{P}$ reference.\cite{BrukerNMRTable} 
The total uncertainty of the field calibration is estimated as $0$ to $\nwu{150}{ppm}$ (mostly correlated and systematic), with a tendency to overestimate the magnetic field at the sample \mpc{position}. The spin-lattice relaxation rate $T_1^{-1}$ was obtained by fitting a stretched-exponential recovery, \mpc{with} stretching exponents ranging from $0.6$ to $1$. The single-crystalline \BCPO{} sample\cite{Wang2010} was mounted on an NMR probe featuring a two-axis rotator. After pre-alignment in a superconducting magnet \aprc{at $\nwu{15}{T}$} (using the angular dependence\cite{DissCasola} of the $^{31}\chemElement{P}$-NMR shift), the sample was realigned \emph{in-situ} \mpc{in} the \mpc{high-field} magnet.
Differences with respect to Ref.~\citenum{Casola2013} are attributed to the lack of such an alignment facility in the previous experiments. Nevertheless, a small residual misalignment remained due to technical limitations. 
In fact, the fine structure of the NMR spectra shown in Fig.~\ref{fig:spectraT1} \mpc{(cf.~footnote \ref{note:narrowComponentFeatures})} was found to disappear upon a subtle change of sample orientation (checked at $\mu_0\,H=\nwu{37.7}{T}$).
The low-frequency shoulder visible in the \mpc{dark-blue} data plotted in Fig.~\refpart{fig:spectraT1}{a} appeared to be affected by the sample orientation as well, which suggests an experimental origin. Additional $^{31}\chemElement{P}$-NMR and magnetization measurements performed after the high-field experiments confirmed the \mpc{sample integrity}.
}

\subsection{High-field NMR: Relative intensities.}
\footnotesize{%
The relative intensity of the two double-horn components fitted to the data in Fig.~\refpart{fig:spectraT1}{c} is approximately $3 : 2$. \mpc{Yet}, the \mpc{group of reflections} linking the four translationally-inequivalent $^{31}\chemElement{P}$ sites (see main text) is isomorphic to $Z_2 \times Z_2$. Therefore, essential degeneracies can only give rise to one, two, or four double-horn \mpc{contributions} with \emph{equal} intensities.
There are two \mpc{possible} causes for the apparently reduced amplitude of the narrow component of the spectrum.
First, the nuclear-spin dynamics are \mpc{most likely non-uniform} across the NMR spectrum, as evidenced by the fact that a comb of 50 saturation pulses \mpc{fully suppressed} the signal at frequencies close to the edge of the spectrum\mpc{,} while failing to do so for frequencies near the center of the \mpc{line} (at $\mu_0{}\,H = \nwu{42.2}{T}$, $T = \nwu{1.7(2)}{K}$; nuclear polarization probed after $\nwu{30}{ms} \ll T_1$). 
Second, the NMR spectra were obtained by summing Fourier-transformed spin-echoes recorded at different frequencies.\cite{Clark1995}
Since signals were strong in general, non-linear response of the receiving electronics could have led to reduced intensities of the central peak.
}

\subsection{Coupling parameters and uncertainties.}
\footnotesize{%
The hyperfine field $\vec{B}_\text{hf}$ at each $^{31}\chemElement{P}$ nucleus is written as $\vec{B}_\text{hf} = \sum_i (A_i + D_i)\, \vec{\mu}_i$, with the sum running over all electronic magnetic moments $\vec{\mu}_i = -\mu_{\mathrm{B}}\,g_i\,\langle \vec{S}_i \rangle$.
After fixing the exchange couplings $(J_1,J_2^\prime,J_2,J_\perp)$ proposed in Ref.~\citenum{Tsirlin2010}, the g-tensors $g_i$ of the two magnetic sites \mpc{were} obtained by fitting the results of full-spectrum exact-diagonalization calculations to the low-field magnetization data from Ref.~\citenum{DissCasola}.\cite{DissPikulski}
Approximate $C_{2v}$-symmetry\mpcc{\cite{DissCasola}} of the $\chemElement{CuO}_4$ plaquettes\cite{Abraham1994} is assumed. We further assume that the g-tensors are symmetric and positive. It turns out that only one of the two approximate $C_{2v}$-symmetry--axes differs significantly between the two $\chemElement{Cu}$ sites. 
Subsequently, the anisotropic hyperfine couplings $A_i$\mpc{,} linking\cite{Alexander2010} each $\chemElement{P}$ nucleus to four surrounding magnetic sites [Fig.~\refpart{fig:structure}{b}]\mpc{,} can be estimated from the angular dependence of the NMR shift in the paramagnetic state of pristine and slightly $\chemElement{Zn}$-doped \BCPO{}.\cite{DissCasola,DissPikulski}
We assume that the corresponding matrices are symmetric.\mpcc{\cite{DissCasola}}
The dipole couplings $D_i$ are treated \mpc{by means of a plane-wise summation} technique,\cite{Nijboer1957,DeWette1965} \mpc{using} the crystal structure \mpc{reported in} Ref.~\citenum{Abraham1994}.
By repeating the analysis with $J_2^\prime = J_2 = 0.75\,J_1$, we estimate uncertainties up to the order of $25\%$ for the elements of the matrices $A_i\,g_i$, and $10\%$ for \mpc{those} of $g_i$.\cite{DissPikulski} 
Note, however, that more pessimistic considerations suggest errors of order $100\%$ and more.\cite{DissPikulski}
\mpc{The details of the above analyses are beyond the scope of this work and have been documented in Refs.~\citenum{DissCasola} and \citenum{DissPikulski}.}
}

\subsection{Models for \mpc{the} NMR line shape.}
\footnotesize{%

The NMR spectrum expected for any hypothetical magnetic structure can be calculated if the following properties are known: \begin{inparaenum}[(i)] \item the g-tensors for both magnetic sites, \item the matrices describing the hyperfine couplings, and \item the crystal structure\end{inparaenum}.
While the former two can be estimated from low-field data, the resulting parameters have large uncertainties \mpc{(see preceding section)}.
\mpc{Moreover, a magnetic field lowers the crystal symmetry (see main text), such that field-induced lattice} deformations could \mpc{in principle} alter any of the three aforementioned properties.
\mpc{In addition}, the quantitative information contained in the NMR spectrum [Fig.~\refpart{fig:spectraT1}{c}]---i.\,e.\,, the widths and the center frequency of the two double-horn components---is insufficient to constrain even the simplest spiral models.
\mpc{Despite these limitations, quantitative analyses of the NMR line shape were attempted and shall be briefly described in the following.}

Following the main text, a spiral magnetic structure analogous to \mpc{that shown in} Fig.~\ref{fig:tilt} is assumed \mpc{to form} within each magnetic layer of \BCPO{}. The propagation wavenumber along $b$ is fixed to ${q_b = 0.574}$.\PlumbNatPhyscite{Plumb2013}
\mpc{We also include a tilt of the spiral axis, as it is expected to occur due to the recently-suggested\PlumbNatPhyscite{} DM interaction $D_1^{ac}$.}
The twist-distortion associated with the \mpc{also} suggested\PlumbNatPhyscite{} DM interaction $D_1^b$ is not considered---in order to limit the number of degrees of freedom and because numerical calculations suggest that the associated angles are smaller.\cite{DissPikulski}
Since the longitudinal magnetization determines \mpc{only} the NMR shift\mpc{,} but not the \mpc{NMR} line shape, we end up with five parameters (the wavenumber $q_a$ describing the propagation of the magnetic structure along $a$, two tilt angles, and two transverse ordered moments \mpc{$m_{\perp,i} \leq 0.5\,\hbar$})\mpc{,} which are adjusted in order to reproduce the widths of the two observed double-horn contributions shown in Fig.~\refpart{fig:spectraT1}{c}.
\mpc{To} account for the aforementioned uncertainties\mpc{, we consider} corresponding \mpc{one-} and two-parameter variations of the coupling parameters. For each resulting set of coupling parameters, candidate solutions reproducing the experimental data exist.
Furthermore, the aforementioned model has been generalized to include random stacking configurations of \mpc{magnetically-ordered} layers.
In particular, taking a candidate solution with $q_a = 0.5$ and recomputing the line shape assuming a random stacking scenario with co-aligned chirality and phase shift $\delta = \pm\pi/2$ (see main text) yielded only small changes to the spectrum.
\mpc{For additional information, as well as detailed symmetry considerations covering more general spiral magnetic structures, we refer to Ref.~\citenum{DissPikulski}.}
}

\section{Acknowledgements}
\footnotesize{%
\aprc{Work at the National High Magnetic Field Laboratory was supported by the User Collaborative Grants Program (UCGP) under National Science Foundation Cooperative Agreement No. DMR-1157490, and the State of Florida.}
\mpc{The presented calculations made use of the} ALPS libraries and MPS applications\mpc{\cite{Albuquerque2007,Bauer2011,Dolfi2014}} and were executed on the Brutus and Euler clusters of ETH Z\"urich, as well as on Piz Dora at the Swiss National Supercomputing Centre (CSCS). 
M.\,P.\ acknowledges the technical support of M.~Dolfi and useful discussions with \mpc{Ch.}~R\"uegg.
\mpc{The authors also thank A.~E.~Feiguin for sharing a summary of previous unpublished results related to Ref.~\citenum{Casola2013}, according to which the DM-induced distortions considered in detail in this work may already have been noticed in his calculations.
M.\,P. thanks N.~Lavar\'elo, G.~Roux, and N.~Laflorencie for providing the data points of the phase boundaries reported in Ref.~\citenum{Lavarelo2011} \mpcc{(see Fig.~\ref{fig:phaseDiagramChirality})}.}
This work was financially supported in part by the Schweizerische Nationalfonds zur F\"{o}rderung der Wissenschaftlichen Forschung (SNF).
}

\section{Author contributions}
\footnotesize{%
Sample growth: S.\,W.\,; initial proposal: F.\,C.\,; 
experiments: M.\,P.\,, T.\,S.\,, F.\,C.\,, P.\,L.\,K.\,, A.\,P.\,R.\,;
data analysis: M.\,P.\,, based on earlier work by F.\,C.\,; calculations: M.\,P.\,;
manuscript based on draft by M.\,P.\ and contributions from the co-authors, \tsc{in particular T.\,S.\ and H.-R.\,O.}.
}

\footnotesize{%
\vspace{2mm}
\noindent\textbf{Competing financial interests:} The authors declare 
no competing financial interests.
}

\vspace{-5mm}
\renewcommand*{\bibfont}{\footnotesize}

\bibliography{paper}

\begin{thebibliography}{74}%
\makeatletter
\providecommand \@ifxundefined [1]{%
 \@ifx{#1\undefined}
}%
\providecommand \@ifnum [1]{%
 \ifnum #1\expandafter \@firstoftwo
 \else \expandafter \@secondoftwo
 \fi
}%
\providecommand \@ifx [1]{%
 \ifx #1\expandafter \@firstoftwo
 \else \expandafter \@secondoftwo
 \fi
}%
\providecommand \natexlab [1]{#1}%
\providecommand \enquote  [1]{``#1''}%
\providecommand \bibnamefont  [1]{#1}%
\providecommand \bibfnamefont [1]{#1}%
\providecommand \citenamefont [1]{#1}%
\providecommand \href@noop [0]{\@secondoftwo}%
\providecommand \href [0]{\begingroup \@sanitize@url \@href}%
\providecommand \@href[1]{\@@startlink{#1}\@@href}%
\providecommand \@@href[1]{\endgroup#1\@@endlink}%
\providecommand \@sanitize@url [0]{\catcode `\\12\catcode `\$12\catcode
  `\&12\catcode `\#12\catcode `\^12\catcode `\_12\catcode `\%12\relax}%
\providecommand \@@startlink[1]{}%
\providecommand \@@endlink[0]{}%
\providecommand \url  [0]{\begingroup\@sanitize@url \@url }%
\providecommand \@url [1]{\endgroup\@href {#1}{\urlprefix }}%
\providecommand \urlprefix  [0]{URL }%
\providecommand \Eprint [0]{\href }%
\providecommand \doibase [0]{http://dx.doi.org/}%
\providecommand \selectlanguage [0]{\@gobble}%
\providecommand \bibinfo  [0]{\@secondoftwo}%
\providecommand \bibfield  [0]{\@secondoftwo}%
\providecommand \translation [1]{[#1]}%
\providecommand \BibitemOpen [0]{}%
\providecommand \bibitemStop [0]{}%
\providecommand \bibitemNoStop [0]{.\EOS\space}%
\providecommand \EOS [0]{\spacefactor3000\relax}%
\providecommand \BibitemShut  [1]{\csname bibitem#1\endcsname}%
\let\auto@bib@innerbib\@empty
%</preamble>
\bibitem [{\citenamefont {Tsirlin}\ \emph {et~al.}(2010)\citenamefont
  {Tsirlin}, \citenamefont {Rousochatzakis}, \citenamefont {Kasinathan},
  \citenamefont {Janson}, \citenamefont {Nath}, \citenamefont {Weickert},
  \citenamefont {Geibel}, \citenamefont {L\"auchli},\ and\ \citenamefont
  {Rosner}}]{Tsirlin2010}%
  \BibitemOpen
  \bibfield  {author} {\bibinfo {author} {\bibfnamefont {A.~A.}\ \bibnamefont
  {Tsirlin}}, \bibinfo {author} {\bibfnamefont {I.}~\bibnamefont
  {Rousochatzakis}}, \bibinfo {author} {\bibfnamefont {D.}~\bibnamefont
  {Kasinathan}}, \bibinfo {author} {\bibfnamefont {O.}~\bibnamefont {Janson}},
  \bibinfo {author} {\bibfnamefont {R.}~\bibnamefont {Nath}}, \bibinfo {author}
  {\bibfnamefont {F.}~\bibnamefont {Weickert}}, \bibinfo {author}
  {\bibfnamefont {C.}~\bibnamefont {Geibel}}, \bibinfo {author} {\bibfnamefont
  {A.~M.}\ \bibnamefont {L\"auchli}}, \ and\ \bibinfo {author} {\bibfnamefont
  {H.}~\bibnamefont {Rosner}},\ }\bibfield  {title} {\enquote {\bibinfo {title}
  {Bridging frustrated-spin-chain and spin-ladder physics:
  Quasi-one-dimensional magnetism of {$\text{BiCu}_2\text{PO}_6$}},}\ }\href
  {\doibase 10.1103/PhysRevB.82.144426} {\bibfield  {journal} {\bibinfo
  {journal} {Phys. Rev. B}\ }\textbf {\bibinfo {volume} {82}},\ \bibinfo
  {pages} {144426} (\bibinfo {year} {2010})}\BibitemShut {NoStop}%
\bibitem [{\citenamefont {Shyiko}\ \emph {et~al.}(2013)\citenamefont {Shyiko},
  \citenamefont {McCulloch}, \citenamefont {Gumenjuk-Sichevska},\ and\
  \citenamefont {Kolezhuk}}]{Shyiko2013}%
  \BibitemOpen
  \bibfield  {author} {\bibinfo {author} {\bibfnamefont {I.~T.}\ \bibnamefont
  {Shyiko}}, \bibinfo {author} {\bibfnamefont {I.~P.}\ \bibnamefont
  {McCulloch}}, \bibinfo {author} {\bibfnamefont {J.~V.}\ \bibnamefont
  {Gumenjuk-Sichevska}}, \ and\ \bibinfo {author} {\bibfnamefont {A.~K.}\
  \bibnamefont {Kolezhuk}},\ }\bibfield  {title} {\enquote {\bibinfo {title}
  {Double zigzag spin chain in a strong magnetic field close to saturation},}\
  }\href {\doibase 10.1103/PhysRevB.88.014403} {\bibfield  {journal} {\bibinfo
  {journal} {Phys. Rev. B}\ }\textbf {\bibinfo {volume} {88}},\ \bibinfo
  {pages} {014403} (\bibinfo {year} {2013})}\BibitemShut {NoStop}%
\bibitem [{\citenamefont {Sugimoto}\ \emph
  {et~al.}(2015{\natexlab{a}})\citenamefont {Sugimoto}, \citenamefont {Mori},
  \citenamefont {Tohyama},\ and\ \citenamefont
  {Maekawa}}]{Sugimito2014jpsconf}%
  \BibitemOpen
  \bibfield  {author} {\bibinfo {author} {\bibfnamefont {Takanori}\
  \bibnamefont {Sugimoto}}, \bibinfo {author} {\bibfnamefont {Michiyasu}\
  \bibnamefont {Mori}}, \bibinfo {author} {\bibfnamefont {Takami}\ \bibnamefont
  {Tohyama}}, \ and\ \bibinfo {author} {\bibfnamefont {Sadamichi}\ \bibnamefont
  {Maekawa}},\ }\bibfield  {title} {\enquote {\bibinfo {title} {Lifshitz
  {T}ransition {I}nduced by {M}agnetic {F}ield in {F}rustrated {T}wo-{L}eg
  {S}pin-{L}adder {S}ystems},}\ }in\ \href {\doibase 10.7566/JPSCP.8.034005}
  {\emph {\bibinfo {booktitle} {Proc. 2nd {I}nt. {S}ymp. {S}cience at {J-PARC}
  --- {U}nlocking the {M}ysteries of {L}ife, {M}atter and the {U}niverse ---,
  {T}sukuba, {I}baraki, {J}apan}}},\ \bibinfo {series and number} {\bibinfo
  {series} {JPS Conf. Proc.}\ No.\ \bibinfo {number} {8, 034005}}\ (\bibinfo
  {year} {2015})\BibitemShut {NoStop}%
\bibitem [{\citenamefont {Sugimoto}\ \emph
  {et~al.}(2015{\natexlab{b}})\citenamefont {Sugimoto}, \citenamefont {Mori},
  \citenamefont {Tohyama},\ and\ \citenamefont {Maekawa}}]{Sugimoto2015}%
  \BibitemOpen
  \bibfield  {author} {\bibinfo {author} {\bibfnamefont {Takanori}\
  \bibnamefont {Sugimoto}}, \bibinfo {author} {\bibfnamefont {Michiyasu}\
  \bibnamefont {Mori}}, \bibinfo {author} {\bibfnamefont {Takami}\ \bibnamefont
  {Tohyama}}, \ and\ \bibinfo {author} {\bibfnamefont {Sadamichi}\ \bibnamefont
  {Maekawa}},\ }\bibfield  {title} {\enquote {\bibinfo {title} {Magnetization
  plateaus by reconstructed quasispinons in a frustrated two-leg spin ladder
  under a magnetic field},}\ }\href {\doibase 10.1103/PhysRevB.92.125114}
  {\bibfield  {journal} {\bibinfo  {journal} {Phys. Rev. B}\ }\textbf {\bibinfo
  {volume} {92}},\ \bibinfo {pages} {125114} (\bibinfo {year}
  {2015}{\natexlab{b}})}\BibitemShut {NoStop}%
\bibitem [{\citenamefont {Casola}\ \emph {et~al.}(2013)\citenamefont {Casola},
  \citenamefont {Shiroka}, \citenamefont {Feiguin}, \citenamefont {Wang},
  \citenamefont {Grbi\'{c}}, \citenamefont {Horvati\'{c}}, \citenamefont
  {Kr\"amer}, \citenamefont {Mukhopadhyay}, \citenamefont {Conder},
  \citenamefont {Berthier}, \citenamefont {Ott}, \citenamefont {R\o{}nnow},
  \citenamefont {R\"uegg},\ and\ \citenamefont {Mesot}}]{Casola2013}%
  \BibitemOpen
  \bibfield  {author} {\bibinfo {author} {\bibfnamefont {F.}~\bibnamefont
  {Casola}}, \bibinfo {author} {\bibfnamefont {T.}~\bibnamefont {Shiroka}},
  \bibinfo {author} {\bibfnamefont {A.}~\bibnamefont {Feiguin}}, \bibinfo
  {author} {\bibfnamefont {S.}~\bibnamefont {Wang}}, \bibinfo {author}
  {\bibfnamefont {M.~S.}\ \bibnamefont {Grbi\'{c}}}, \bibinfo {author}
  {\bibfnamefont {M.}~\bibnamefont {Horvati\'{c}}}, \bibinfo {author}
  {\bibfnamefont {S.}~\bibnamefont {Kr\"amer}}, \bibinfo {author}
  {\bibfnamefont {S.}~\bibnamefont {Mukhopadhyay}}, \bibinfo {author}
  {\bibfnamefont {K.}~\bibnamefont {Conder}}, \bibinfo {author} {\bibfnamefont
  {C.}~\bibnamefont {Berthier}}, \bibinfo {author} {\bibfnamefont {H.-R.}\
  \bibnamefont {Ott}}, \bibinfo {author} {\bibfnamefont {H.~M.}\ \bibnamefont
  {R\o{}nnow}}, \bibinfo {author} {\bibfnamefont {{\relax Ch}.}~\bibnamefont
  {R\"uegg}}, \ and\ \bibinfo {author} {\bibfnamefont {J.}~\bibnamefont
  {Mesot}},\ }\bibfield  {title} {\enquote {\bibinfo {title} {Field-{I}nduced
  {Q}uantum {S}oliton {L}attice in a {F}rustrated {T}wo-{L}eg {S}pin-{$1/2$}
  {L}adder},}\ }\href {\doibase 10.1103/PhysRevLett.110.187201} {\bibfield
  {journal} {\bibinfo  {journal} {Phys. Rev. Lett.}\ }\textbf {\bibinfo
  {volume} {110}},\ \bibinfo {pages} {187201} (\bibinfo {year}
  {2013})}\BibitemShut {NoStop}%
\bibitem [{\citenamefont {Casola}\ \emph {et~al.}(2012)\citenamefont {Casola},
  \citenamefont {Shiroka}, \citenamefont {Feiguin}, \citenamefont {Wang},
  \citenamefont {Grbi\'{c}}, \citenamefont {Horvati\'{c}}, \citenamefont
  {Kr\"amer}, \citenamefont {Mukhopadhyay}, \citenamefont {Berthier},
  \citenamefont {Ott}, \citenamefont {R\o{}nnow}, \citenamefont {R\"uegg},\
  and\ \citenamefont {Mesot}}]{Casola2013arXiv}%
  \BibitemOpen
  \bibfield  {author} {\bibinfo {author} {\bibfnamefont {F.}~\bibnamefont
  {Casola}}, \bibinfo {author} {\bibfnamefont {T.}~\bibnamefont {Shiroka}},
  \bibinfo {author} {\bibfnamefont {A.}~\bibnamefont {Feiguin}}, \bibinfo
  {author} {\bibfnamefont {S.}~\bibnamefont {Wang}}, \bibinfo {author}
  {\bibfnamefont {M.~S.}\ \bibnamefont {Grbi\'{c}}}, \bibinfo {author}
  {\bibfnamefont {M.}~\bibnamefont {Horvati\'{c}}}, \bibinfo {author}
  {\bibfnamefont {S.}~\bibnamefont {Kr\"amer}}, \bibinfo {author}
  {\bibfnamefont {S.}~\bibnamefont {Mukhopadhyay}}, \bibinfo {author}
  {\bibfnamefont {C.}~\bibnamefont {Berthier}}, \bibinfo {author}
  {\bibfnamefont {H.-R.}\ \bibnamefont {Ott}}, \bibinfo {author} {\bibfnamefont
  {H.~M.}\ \bibnamefont {R\o{}nnow}}, \bibinfo {author} {\bibfnamefont {Ch.}\
  \bibnamefont {R\"uegg}}, \ and\ \bibinfo {author} {\bibfnamefont
  {J.}~\bibnamefont {Mesot}},\ }\bibfield  {title} {\enquote {\bibinfo {title}
  {Field-induced quantum soliton lattice in a frustrated two-leg spin-1/2
  ladder},}\ }\href {https://arxiv.org/abs/1211.5522v1} {\bibfield  {journal}
  {\bibinfo  {journal} {arXiv e-prints}\ } (\bibinfo {year} {2012})},\ \Eprint
  {http://arxiv.org/abs/arXiv:1211.5522v1} {arXiv:1211.5522v1} \BibitemShut
  {NoStop}%
\bibitem [{\citenamefont {Vekua}\ and\ \citenamefont
  {Honecker}(2006)}]{Vekua2006}%
  \BibitemOpen
  \bibfield  {author} {\bibinfo {author} {\bibfnamefont {T.}~\bibnamefont
  {Vekua}}\ and\ \bibinfo {author} {\bibfnamefont {A.}~\bibnamefont
  {Honecker}},\ }\bibfield  {title} {\enquote {\bibinfo {title} {Quantum dimer
  phases in a frustrated spin ladder: Effective field theory approach and exact
  diagonalization},}\ }\href {\doibase 10.1103/PhysRevB.73.214427} {\bibfield
  {journal} {\bibinfo  {journal} {Phys. Rev. B}\ }\textbf {\bibinfo {volume}
  {73}},\ \bibinfo {pages} {214427} (\bibinfo {year} {2006})}\BibitemShut
  {NoStop}%
\bibitem [{\citenamefont {Dagotto}\ and\ \citenamefont
  {Rice}(1996)}]{Dagotto1996}%
  \BibitemOpen
  \bibfield  {author} {\bibinfo {author} {\bibfnamefont {E.}~\bibnamefont
  {Dagotto}}\ and\ \bibinfo {author} {\bibfnamefont {T.~M.}\ \bibnamefont
  {Rice}},\ }\bibfield  {title} {\enquote {\bibinfo {title} {Surprises on the
  {W}ay from {O}ne- to {T}wo-{D}imensional {Q}uantum {M}agnets: {T}he {L}adder
  {M}aterials},}\ }\href {\doibase 10.1126/science.271.5249.618} {\bibfield
  {journal} {\bibinfo  {journal} {Science}\ }\textbf {\bibinfo {volume}
  {271}},\ \bibinfo {pages} {618--623} (\bibinfo {year} {1996})}\BibitemShut
  {NoStop}%
\bibitem [{\citenamefont {Majumdar}\ and\ \citenamefont
  {Ghosh}(1969{\natexlab{a}})}]{Majumdar1967_1}%
  \BibitemOpen
  \bibfield  {author} {\bibinfo {author} {\bibfnamefont {C.~K.}\ \bibnamefont
  {Majumdar}}\ and\ \bibinfo {author} {\bibfnamefont {D.~K.}\ \bibnamefont
  {Ghosh}},\ }\bibfield  {title} {\enquote {\bibinfo {title} {On
  {N}ext-{N}earest-{N}eighbor {I}nteraction in {L}inear {C}hain. {I}},}\ }\href
  {\doibase 10.1063/1.1664978} {\bibfield  {journal} {\bibinfo  {journal} {J.
  Math. Phys.}\ }\textbf {\bibinfo {volume} {10}},\ \bibinfo {pages}
  {1388--1398} (\bibinfo {year} {1969}{\natexlab{a}})}\BibitemShut {NoStop}%
\bibitem [{\citenamefont {Majumdar}\ and\ \citenamefont
  {Ghosh}(1969{\natexlab{b}})}]{Majumdar1967_2}%
  \BibitemOpen
  \bibfield  {author} {\bibinfo {author} {\bibfnamefont {C.~K.}\ \bibnamefont
  {Majumdar}}\ and\ \bibinfo {author} {\bibfnamefont {D.~K.}\ \bibnamefont
  {Ghosh}},\ }\bibfield  {title} {\enquote {\bibinfo {title} {On
  {N}ext-{N}earest-{N}eighbor {I}nteraction in {L}inear {C}hain. {II}},}\
  }\href {\doibase 10.1063/1.1664979} {\bibfield  {journal} {\bibinfo
  {journal} {J. Math. Phys.}\ }\textbf {\bibinfo {volume} {10}},\ \bibinfo
  {pages} {1399--1402} (\bibinfo {year} {1969}{\natexlab{b}})}\BibitemShut
  {NoStop}%
\bibitem [{\citenamefont {Lavar\'elo}\ \emph {et~al.}(2011)\citenamefont
  {Lavar\'elo}, \citenamefont {Roux},\ and\ \citenamefont
  {Laflorencie}}]{Lavarelo2011}%
  \BibitemOpen
  \bibfield  {author} {\bibinfo {author} {\bibfnamefont {A.}~\bibnamefont
  {Lavar\'elo}}, \bibinfo {author} {\bibfnamefont {G.}~\bibnamefont {Roux}}, \
  and\ \bibinfo {author} {\bibfnamefont {N.}~\bibnamefont {Laflorencie}},\
  }\bibfield  {title} {\enquote {\bibinfo {title} {Melting of a
  frustration-induced dimer crystal and incommensurability in the
  {${J}_{1}$-${J}_{2}$} two-leg ladder},}\ }\href {\doibase
  10.1103/PhysRevB.84.144407} {\bibfield  {journal} {\bibinfo  {journal} {Phys.
  Rev. B}\ }\textbf {\bibinfo {volume} {84}},\ \bibinfo {pages} {144407}
  (\bibinfo {year} {2011})}\BibitemShut {NoStop}%
\bibitem [{\citenamefont {Shastry}\ and\ \citenamefont
  {Sutherland}(1981)}]{Shastry1981}%
  \BibitemOpen
  \bibfield  {author} {\bibinfo {author} {\bibfnamefont {B.~S.}\ \bibnamefont
  {Shastry}}\ and\ \bibinfo {author} {\bibfnamefont {B.}~\bibnamefont
  {Sutherland}},\ }\bibfield  {title} {\enquote {\bibinfo {title} {Excitation
  {S}pectrum of a {D}imerized {N}ext-{N}eighbor {A}ntiferromagnetic {C}hain},}\
  }\href {\doibase 10.1103/PhysRevLett.47.964} {\bibfield  {journal} {\bibinfo
  {journal} {Phys. Rev. Lett.}\ }\textbf {\bibinfo {volume} {47}},\ \bibinfo
  {pages} {964--967} (\bibinfo {year} {1981})}\BibitemShut {NoStop}%
\bibitem [{\citenamefont {Bursill}\ \emph {et~al.}(1995)\citenamefont
  {Bursill}, \citenamefont {Gehring}, \citenamefont {Farnell}, \citenamefont
  {Parkinson}, \citenamefont {Xiang},\ and\ \citenamefont
  {Zeng}}]{Bursill1995}%
  \BibitemOpen
  \bibfield  {author} {\bibinfo {author} {\bibfnamefont {R.}~\bibnamefont
  {Bursill}}, \bibinfo {author} {\bibfnamefont {G.~A.}\ \bibnamefont
  {Gehring}}, \bibinfo {author} {\bibfnamefont {D.~J.~J.}\ \bibnamefont
  {Farnell}}, \bibinfo {author} {\bibfnamefont {J.~B.}\ \bibnamefont
  {Parkinson}}, \bibinfo {author} {\bibfnamefont {T.}~\bibnamefont {Xiang}}, \
  and\ \bibinfo {author} {\bibfnamefont {C.}~\bibnamefont {Zeng}},\ }\bibfield
  {title} {\enquote {\bibinfo {title} {Numerical and approximate analytical
  results for the frustrated spin-{$\frac{1}{2}$} quantum spin chain},}\ }\href
  {\doibase 10.1088/0953-8984/7/45/016} {\bibfield  {journal} {\bibinfo
  {journal} {J. Phys.: Condens. Matter}\ }\textbf {\bibinfo {volume} {7}},\
  \bibinfo {pages} {8605--8618} (\bibinfo {year} {1995})}\BibitemShut {NoStop}%
\bibitem [{\citenamefont {Balents}(2010)}]{Balents2010}%
  \BibitemOpen
  \bibfield  {author} {\bibinfo {author} {\bibfnamefont {L.}~\bibnamefont
  {Balents}},\ }\bibfield  {title} {\enquote {\bibinfo {title} {Spin liquids in
  frustrated magnets},}\ }\href {\doibase 10.1038/nature08917} {\bibfield
  {journal} {\bibinfo  {journal} {Nature}\ }\textbf {\bibinfo {volume} {464}},\
  \bibinfo {pages} {199--208} (\bibinfo {year} {2010})}\BibitemShut {NoStop}%
\bibitem [{\citenamefont {Wessel}\ and\ \citenamefont
  {Haas}(2000)}]{Wessel2000}%
  \BibitemOpen
  \bibfield  {author} {\bibinfo {author} {\bibfnamefont {S.}~\bibnamefont
  {Wessel}}\ and\ \bibinfo {author} {\bibfnamefont {S.}~\bibnamefont {Haas}},\
  }\bibfield  {title} {\enquote {\bibinfo {title} {Magnetic field induced
  ordering in quasi-one-dimensional quantum magnets},}\ }\href {\doibase
  10.1007/s100510070194} {\bibfield  {journal} {\bibinfo  {journal} {Eur. Phys.
  J. B}\ }\textbf {\bibinfo {volume} {16}},\ \bibinfo {pages} {393--396}
  (\bibinfo {year} {2000})}\BibitemShut {NoStop}%
\bibitem [{\citenamefont {Giamarchi}\ \emph {et~al.}(2008)\citenamefont
  {Giamarchi}, \citenamefont {R\"uegg},\ and\ \citenamefont
  {Tchernyshyov}}]{Giamarchi2008}%
  \BibitemOpen
  \bibfield  {author} {\bibinfo {author} {\bibfnamefont {T.}~\bibnamefont
  {Giamarchi}}, \bibinfo {author} {\bibfnamefont {{\relax Ch}.}~\bibnamefont
  {R\"uegg}}, \ and\ \bibinfo {author} {\bibfnamefont {O.}~\bibnamefont
  {Tchernyshyov}},\ }\bibfield  {title} {\enquote {\bibinfo {title}
  {{B}ose-{E}instein condensation in magnetic insulators},}\ }\href {\doibase
  10.1038/nphys893} {\bibfield  {journal} {\bibinfo  {journal} {Nat. Phys.}\
  }\textbf {\bibinfo {volume} {4}},\ \bibinfo {pages} {198--204} (\bibinfo
  {year} {2008})}\BibitemShut {NoStop}%
\bibitem [{\citenamefont {Nomura}\ and\ \citenamefont
  {Okamoto}(1994)}]{Nomura1994}%
  \BibitemOpen
  \bibfield  {author} {\bibinfo {author} {\bibfnamefont {K.}~\bibnamefont
  {Nomura}}\ and\ \bibinfo {author} {\bibfnamefont {K.}~\bibnamefont
  {Okamoto}},\ }\bibfield  {title} {\enquote {\bibinfo {title} {Critical
  properties of {$S= 1/2$} antiferromagnetic {$XXZ$} chain with
  next-nearest-neighbour interactions},}\ }\href {\doibase
  10.1088/0305-4470/27/17/012} {\bibfield  {journal} {\bibinfo  {journal} {J.
  Phys. A: Math. Gen.}\ }\textbf {\bibinfo {volume} {27}},\ \bibinfo {pages}
  {5773--5788} (\bibinfo {year} {1994})}\BibitemShut {NoStop}%
\bibitem [{\citenamefont {Chitra}\ \emph {et~al.}(1995)\citenamefont {Chitra},
  \citenamefont {Pati}, \citenamefont {Krishnamurthy}, \citenamefont {Sen},\
  and\ \citenamefont {Ramasesha}}]{Chitra1995}%
  \BibitemOpen
  \bibfield  {author} {\bibinfo {author} {\bibfnamefont {R.}~\bibnamefont
  {Chitra}}, \bibinfo {author} {\bibfnamefont {S.}~\bibnamefont {Pati}},
  \bibinfo {author} {\bibfnamefont {H.~R.}\ \bibnamefont {Krishnamurthy}},
  \bibinfo {author} {\bibfnamefont {D.}~\bibnamefont {Sen}}, \ and\ \bibinfo
  {author} {\bibfnamefont {S.}~\bibnamefont {Ramasesha}},\ }\bibfield  {title}
  {\enquote {\bibinfo {title} {Density-matrix renormalization-group studies of
  the spin-1/2 {H}eisenberg system with dimerization and frustration},}\ }\href
  {\doibase 10.1103/PhysRevB.52.6581} {\bibfield  {journal} {\bibinfo
  {journal} {Phys. Rev. B}\ }\textbf {\bibinfo {volume} {52}},\ \bibinfo
  {pages} {6581--6587} (\bibinfo {year} {1995})}\BibitemShut {NoStop}%
\bibitem [{\citenamefont {Nersesyan}\ \emph {et~al.}(1998)\citenamefont
  {Nersesyan}, \citenamefont {Gogolin},\ and\ \citenamefont
  {E\ss{}ler}}]{Nersesyan1998}%
  \BibitemOpen
  \bibfield  {author} {\bibinfo {author} {\bibfnamefont {A.~A.}\ \bibnamefont
  {Nersesyan}}, \bibinfo {author} {\bibfnamefont {A.~O.}\ \bibnamefont
  {Gogolin}}, \ and\ \bibinfo {author} {\bibfnamefont {F.~H.~L.}\ \bibnamefont
  {E\ss{}ler}},\ }\bibfield  {title} {\enquote {\bibinfo {title}
  {Incommensurate {S}pin {C}orrelations in {S}pin-$1/2$ {F}rustrated
  {T}wo-{L}eg {H}eisenberg {L}adders},}\ }\href {\doibase
  10.1103/PhysRevLett.81.910} {\bibfield  {journal} {\bibinfo  {journal} {Phys.
  Rev. Lett.}\ }\textbf {\bibinfo {volume} {81}},\ \bibinfo {pages} {910--913}
  (\bibinfo {year} {1998})}\BibitemShut {NoStop}%
\bibitem [{\citenamefont {Hikihara}\ \emph {et~al.}(2010)\citenamefont
  {Hikihara}, \citenamefont {Momoi}, \citenamefont {Furusaki},\ and\
  \citenamefont {Kawamura}}]{Hikihara2010}%
  \BibitemOpen
  \bibfield  {author} {\bibinfo {author} {\bibfnamefont {T.}~\bibnamefont
  {Hikihara}}, \bibinfo {author} {\bibfnamefont {T.}~\bibnamefont {Momoi}},
  \bibinfo {author} {\bibfnamefont {A.}~\bibnamefont {Furusaki}}, \ and\
  \bibinfo {author} {\bibfnamefont {H.}~\bibnamefont {Kawamura}},\ }\bibfield
  {title} {\enquote {\bibinfo {title} {Magnetic phase diagram of the spin-1/2
  antiferromagnetic zigzag ladder},}\ }\href {\doibase
  10.1103/PhysRevB.81.224433} {\bibfield  {journal} {\bibinfo  {journal} {Phys.
  Rev. B}\ }\textbf {\bibinfo {volume} {81}},\ \bibinfo {pages} {224433}
  (\bibinfo {year} {2010})}\BibitemShut {NoStop}%
\bibitem [{\citenamefont {Okunishi}(2008)}]{Okunishi2008}%
  \BibitemOpen
  \bibfield  {author} {\bibinfo {author} {\bibfnamefont {K.}~\bibnamefont
  {Okunishi}},\ }\bibfield  {title} {\enquote {\bibinfo {title} {On
  {C}alculation of {V}ector {S}pin {C}hirality for {Z}igzag {S}pin {C}hains},}\
  }\href {\doibase 10.1143/JPSJ.77.114004} {\bibfield  {journal} {\bibinfo
  {journal} {J. Phys. Soc. Jpn.}\ }\textbf {\bibinfo {volume} {77}},\ \bibinfo
  {pages} {114004} (\bibinfo {year} {2008})}\BibitemShut {NoStop}%
\bibitem [{\citenamefont {Kolezhuk}\ and\ \citenamefont
  {Vekua}(2005)}]{Kolezhuk2005}%
  \BibitemOpen
  \bibfield  {author} {\bibinfo {author} {\bibfnamefont {A.}~\bibnamefont
  {Kolezhuk}}\ and\ \bibinfo {author} {\bibfnamefont {T.}~\bibnamefont
  {Vekua}},\ }\bibfield  {title} {\enquote {\bibinfo {title} {Field-induced
  chiral phase in isotropic frustrated spin chains},}\ }\href {\doibase
  10.1103/PhysRevB.72.094424} {\bibfield  {journal} {\bibinfo  {journal} {Phys.
  Rev. B}\ }\textbf {\bibinfo {volume} {72}},\ \bibinfo {pages} {094424}
  (\bibinfo {year} {2005})}\BibitemShut {NoStop}%
\bibitem [{\citenamefont {McCulloch}\ \emph {et~al.}(2008)\citenamefont
  {McCulloch}, \citenamefont {Kube}, \citenamefont {Kurz}, \citenamefont
  {Kleine}, \citenamefont {Schollw\"ock},\ and\ \citenamefont
  {Kolezhuk}}]{McCulloch2008}%
  \BibitemOpen
  \bibfield  {author} {\bibinfo {author} {\bibfnamefont {I.~P.}\ \bibnamefont
  {McCulloch}}, \bibinfo {author} {\bibfnamefont {R.}~\bibnamefont {Kube}},
  \bibinfo {author} {\bibfnamefont {M.}~\bibnamefont {Kurz}}, \bibinfo {author}
  {\bibfnamefont {A.}~\bibnamefont {Kleine}}, \bibinfo {author} {\bibfnamefont
  {U.}~\bibnamefont {Schollw\"ock}}, \ and\ \bibinfo {author} {\bibfnamefont
  {A.~K.}\ \bibnamefont {Kolezhuk}},\ }\bibfield  {title} {\enquote {\bibinfo
  {title} {Vector chiral order in frustrated spin chains},}\ }\href {\doibase
  10.1103/PhysRevB.77.094404} {\bibfield  {journal} {\bibinfo  {journal} {Phys.
  Rev. B}\ }\textbf {\bibinfo {volume} {77}},\ \bibinfo {pages} {094404}
  (\bibinfo {year} {2008})}\BibitemShut {NoStop}%
\bibitem [{\citenamefont {Ueda}\ and\ \citenamefont
  {Totsuka}(2009)}]{Ueda2009}%
  \BibitemOpen
  \bibfield  {author} {\bibinfo {author} {\bibfnamefont {H.~T.}\ \bibnamefont
  {Ueda}}\ and\ \bibinfo {author} {\bibfnamefont {K.}~\bibnamefont {Totsuka}},\
  }\bibfield  {title} {\enquote {\bibinfo {title} {Magnon {B}ose-{E}instein
  condensation and various phases of three-dimensonal quantum helimagnets under
  high magnetic field},}\ }\href {\doibase 10.1103/PhysRevB.80.014417}
  {\bibfield  {journal} {\bibinfo  {journal} {Phys. Rev. B}\ }\textbf {\bibinfo
  {volume} {80}},\ \bibinfo {pages} {014417} (\bibinfo {year}
  {2009})}\BibitemShut {NoStop}%
\bibitem [{\citenamefont {Villain}(1977)}]{Villain1977}%
  \BibitemOpen
  \bibfield  {author} {\bibinfo {author} {\bibfnamefont {J.}~\bibnamefont
  {Villain}},\ }\bibfield  {title} {\enquote {\bibinfo {title} {A magnetic
  analogue of stereoisomerism : application to helimagnetism in two
  dimensions},}\ }\href {\doibase 10.1051/jphys:01977003804038500} {\bibfield
  {journal} {\bibinfo  {journal} {J. Phys. France}\ }\textbf {\bibinfo {volume}
  {38}},\ \bibinfo {pages} {385--391} (\bibinfo {year} {1977})}\BibitemShut
  {NoStop}%
\bibitem [{\citenamefont {Villain}(1978)}]{Villain1978}%
  \BibitemOpen
  \bibfield  {author} {\bibinfo {author} {\bibfnamefont {J.}~\bibnamefont
  {Villain}},\ }\bibfield  {title} {\enquote {\bibinfo {title} {Chiral {O}rder
  in {H}elimagnets},}\ }in\ \href@noop {} {\emph {\bibinfo {booktitle}
  {{P}roceedings of the 13th {IUPAP} {C}onference on {S}tatistical {P}hysics,
  {T}echnion-{I}srael {I}nstitute of {T}echnology, {H}aifa, {I}srael}}},\
  \bibinfo {series} {Ann. Isr. Phys. Soc.}, Vol.~\bibinfo {volume} {2},\
  \bibinfo {editor} {edited by\ \bibinfo {editor} {\bibfnamefont
  {D.}~\bibnamefont {Cabib}}, \bibinfo {editor} {\bibfnamefont {C.~G.}\
  \bibnamefont {Kuper}}, \ and\ \bibinfo {editor} {\bibfnamefont
  {I.}~\bibnamefont {Riess}}}\ (\bibinfo {year} {1978})\BibitemShut {NoStop}%
\bibitem [{\citenamefont {Abraham}\ \emph {et~al.}(1994)\citenamefont
  {Abraham}, \citenamefont {Ketatni}, \citenamefont {Mairesse},\ and\
  \citenamefont {Mernari}}]{Abraham1994}%
  \BibitemOpen
  \bibfield  {author} {\bibinfo {author} {\bibfnamefont {F.}~\bibnamefont
  {Abraham}}, \bibinfo {author} {\bibfnamefont {M.}~\bibnamefont {Ketatni}},
  \bibinfo {author} {\bibfnamefont {G.}~\bibnamefont {Mairesse}}, \ and\
  \bibinfo {author} {\bibfnamefont {B.}~\bibnamefont {Mernari}},\ }\bibfield
  {title} {\enquote {\bibinfo {title} {Crystal structure of a new bismuth
  copper oxyphosphate: {$\text{BiCu}_2\text{PO}_6$}},}\ }\href {\doibase
  10.1002/chin.199448018} {\bibfield  {journal} {\bibinfo  {journal} {Eur. J.
  Solid State Inorg. Chem.}\ }\textbf {\bibinfo {volume} {31}},\ \bibinfo
  {pages} {313--323} (\bibinfo {year} {1994})}\BibitemShut {NoStop}%
\bibitem [{\citenamefont {Alexander}\ \emph {et~al.}(2010)\citenamefont
  {Alexander}, \citenamefont {Bobroff}, \citenamefont {Mahajan}, \citenamefont
  {Koteswararao}, \citenamefont {Laflorencie},\ and\ \citenamefont
  {Alet}}]{Alexander2010}%
  \BibitemOpen
  \bibfield  {author} {\bibinfo {author} {\bibfnamefont {L.~K.}\ \bibnamefont
  {Alexander}}, \bibinfo {author} {\bibfnamefont {J.}~\bibnamefont {Bobroff}},
  \bibinfo {author} {\bibfnamefont {A.~V.}\ \bibnamefont {Mahajan}}, \bibinfo
  {author} {\bibfnamefont {B.}~\bibnamefont {Koteswararao}}, \bibinfo {author}
  {\bibfnamefont {N.}~\bibnamefont {Laflorencie}}, \ and\ \bibinfo {author}
  {\bibfnamefont {F.}~\bibnamefont {Alet}},\ }\bibfield  {title} {\enquote
  {\bibinfo {title} {Impurity effects in coupled-ladder
  {${\text{BiCu}}_{2}{\text{PO}}_{6}$} studied by {NMR} and quantum {M}onte
  {C}arlo simulations},}\ }\href {\doibase 10.1103/PhysRevB.81.054438}
  {\bibfield  {journal} {\bibinfo  {journal} {Phys. Rev. B}\ }\textbf {\bibinfo
  {volume} {81}},\ \bibinfo {pages} {054438} (\bibinfo {year}
  {2010})}\BibitemShut {NoStop}%
\bibitem [{\citenamefont {Hwang}\ and\ \citenamefont {Kim}(2016)}]{Hwang2016}%
  \BibitemOpen
  \bibfield  {author} {\bibinfo {author} {\bibfnamefont {K.}~\bibnamefont
  {Hwang}}\ and\ \bibinfo {author} {\bibfnamefont {Y.~B.}\ \bibnamefont
  {Kim}},\ }\bibfield  {title} {\enquote {\bibinfo {title} {Theory of triplon
  dynamics in the quantum magnet {${\mathrm{BiCu}}_{2}{\mathrm{PO}}_{6}$}},}\
  }\href {\doibase 10.1103/PhysRevB.93.235130} {\bibfield  {journal} {\bibinfo
  {journal} {Phys. Rev. B}\ }\textbf {\bibinfo {volume} {93}},\ \bibinfo
  {pages} {235130} (\bibinfo {year} {2016})}\BibitemShut {NoStop}%
\bibitem [{\citenamefont {Plumb}\ \emph {et~al.}(2016)\citenamefont {Plumb},
  \citenamefont {Hwang}, \citenamefont {Qiu}, \citenamefont {Harriger},
  \citenamefont {Granroth}, \citenamefont {Kolesnikov}, \citenamefont {Shu},
  \citenamefont {Chou}, \citenamefont {R\"uegg}, \citenamefont {Kim},\ and\
  \citenamefont {Kim}}]{Plumb2016}%
  \BibitemOpen
  \bibfield  {author} {\bibinfo {author} {\bibfnamefont {K.~W.}\ \bibnamefont
  {Plumb}}, \bibinfo {author} {\bibfnamefont {K.}~\bibnamefont {Hwang}},
  \bibinfo {author} {\bibfnamefont {Y.}~\bibnamefont {Qiu}}, \bibinfo {author}
  {\bibfnamefont {L.~W.}\ \bibnamefont {Harriger}}, \bibinfo {author}
  {\bibfnamefont {G.~E.}\ \bibnamefont {Granroth}}, \bibinfo {author}
  {\bibfnamefont {A.~I.}\ \bibnamefont {Kolesnikov}}, \bibinfo {author}
  {\bibfnamefont {G.~J.}\ \bibnamefont {Shu}}, \bibinfo {author} {\bibfnamefont
  {F.~C.}\ \bibnamefont {Chou}}, \bibinfo {author} {\bibfnamefont {{\relax
  Ch}.}~\bibnamefont {R\"uegg}}, \bibinfo {author} {\bibfnamefont {Y.~B.}\
  \bibnamefont {Kim}}, \ and\ \bibinfo {author} {\bibfnamefont {Y.-J.}\
  \bibnamefont {Kim}},\ }\bibfield  {title} {\enquote {\bibinfo {title}
  {Quasiparticle-continuum level repulsion in a quantum magnet},}\ }\href
  {\doibase 10.1038/nphys3566} {\bibfield  {journal} {\bibinfo  {journal} {Nat.
  Phys.}\ }\textbf {\bibinfo {volume} {12}},\ \bibinfo {pages} {224--230}
  (\bibinfo {year} {2016})}\BibitemShut {NoStop}%
\bibitem [{\citenamefont {Plumb}\ \emph {et~al.}(2014)\citenamefont {Plumb},
  \citenamefont {Hwang}, \citenamefont {Qiu}, \citenamefont {Harriger},
  \citenamefont {Granroth}, \citenamefont {Shu}, \citenamefont {Chou},
  \citenamefont {R\"uegg}, \citenamefont {Kim},\ and\ \citenamefont
  {Kim}}]{Plumb2014arXiv}%
  \BibitemOpen
  \bibfield  {author} {\bibinfo {author} {\bibfnamefont {K.~W.}\ \bibnamefont
  {Plumb}}, \bibinfo {author} {\bibfnamefont {K.}~\bibnamefont {Hwang}},
  \bibinfo {author} {\bibfnamefont {Y.}~\bibnamefont {Qiu}}, \bibinfo {author}
  {\bibfnamefont {L.~W.}\ \bibnamefont {Harriger}}, \bibinfo {author}
  {\bibfnamefont {G.~E.}\ \bibnamefont {Granroth}}, \bibinfo {author}
  {\bibfnamefont {G.~J.}\ \bibnamefont {Shu}}, \bibinfo {author} {\bibfnamefont
  {F.~C.}\ \bibnamefont {Chou}}, \bibinfo {author} {\bibfnamefont {{\relax
  Ch}.}~\bibnamefont {R\"uegg}}, \bibinfo {author} {\bibfnamefont {Y.~B.}\
  \bibnamefont {Kim}}, \ and\ \bibinfo {author} {\bibfnamefont {Y.-J.}\
  \bibnamefont {Kim}},\ }\bibfield  {title} {\enquote {\bibinfo {title} {Giant
  {A}nisotropic {I}nteractions in the {C}opper {B}ased {Q}uantum {M}agnet
  {BiCu}$_2${PO}$_6$},}\ }\href@noop {} {\bibfield  {journal} {\bibinfo
  {journal} {e-print}\ } (\bibinfo {year} {2014})},\ \Eprint
  {http://arxiv.org/abs/1408.2528v1} {arXiv:1408.2528v1 [cond-mat.str-el]}
  \BibitemShut {NoStop}%
\bibitem [{\citenamefont {Momma}\ and\ \citenamefont
  {Izumi}(2011)}]{Momma2011}%
  \BibitemOpen
  \bibfield  {author} {\bibinfo {author} {\bibfnamefont {K.}~\bibnamefont
  {Momma}}\ and\ \bibinfo {author} {\bibfnamefont {F.}~\bibnamefont {Izumi}},\
  }\bibfield  {title} {\enquote {\bibinfo {title} {{{\it VESTA 3} for
  three-dimensional visualization of crystal, volumetric and morphology
  data}},}\ }\href {\doibase 10.1107/S0021889811038970} {\bibfield  {journal}
  {\bibinfo  {journal} {J. Appl. Crystallogr.}\ }\textbf {\bibinfo {volume}
  {44}},\ \bibinfo {pages} {1272--1276} (\bibinfo {year} {2011})}\BibitemShut
  {NoStop}%
\bibitem [{\citenamefont {Kohama}\ \emph {et~al.}(2012)\citenamefont {Kohama},
  \citenamefont {Wang}, \citenamefont {Uchida}, \citenamefont {Pr\v{s}a},
  \citenamefont {Zvyagin}, \citenamefont {Skourski}, \citenamefont {McDonald},
  \citenamefont {Balicas}, \citenamefont {R\o{}nnow}, \citenamefont {R\"uegg},\
  and\ \citenamefont {Jaime}}]{Kohama2012}%
  \BibitemOpen
  \bibfield  {author} {\bibinfo {author} {\bibfnamefont {Y.}~\bibnamefont
  {Kohama}}, \bibinfo {author} {\bibfnamefont {S.}~\bibnamefont {Wang}},
  \bibinfo {author} {\bibfnamefont {A.}~\bibnamefont {Uchida}}, \bibinfo
  {author} {\bibfnamefont {K.}~\bibnamefont {Pr\v{s}a}}, \bibinfo {author}
  {\bibfnamefont {S.}~\bibnamefont {Zvyagin}}, \bibinfo {author} {\bibfnamefont
  {Y.}~\bibnamefont {Skourski}}, \bibinfo {author} {\bibfnamefont {R.~D.}\
  \bibnamefont {McDonald}}, \bibinfo {author} {\bibfnamefont {L.}~\bibnamefont
  {Balicas}}, \bibinfo {author} {\bibfnamefont {H.~M.}\ \bibnamefont
  {R\o{}nnow}}, \bibinfo {author} {\bibfnamefont {{\relax Ch}.}~\bibnamefont
  {R\"uegg}}, \ and\ \bibinfo {author} {\bibfnamefont {M.}~\bibnamefont
  {Jaime}},\ }\bibfield  {title} {\enquote {\bibinfo {title} {Anisotropic
  {C}ascade of {F}ield-{I}nduced {P}hase {T}ransitions in the {F}rustrated
  {S}pin-{L}adder {S}ystem {${\mathrm{BiCu}}_{2}{\mathrm{PO}}_{6}$}},}\ }\href
  {\doibase 10.1103/PhysRevLett.109.167204} {\bibfield  {journal} {\bibinfo
  {journal} {Phys. Rev. Lett.}\ }\textbf {\bibinfo {volume} {109}},\ \bibinfo
  {pages} {167204} (\bibinfo {year} {2012})}\BibitemShut {NoStop}%
\bibitem [{\citenamefont {Kohama}\ \emph {et~al.}(2014)\citenamefont {Kohama},
  \citenamefont {Mochidzuki}, \citenamefont {Terashima}, \citenamefont
  {Miyata}, \citenamefont {DeMuer}, \citenamefont {Klein}, \citenamefont
  {Marcenat}, \citenamefont {Dun}, \citenamefont {Zhou}, \citenamefont {Li},
  \citenamefont {Balicas}, \citenamefont {Abe}, \citenamefont {Matsuda},
  \citenamefont {Takeyama}, \citenamefont {Matsuo},\ and\ \citenamefont
  {Kindo}}]{Kohama2014}%
  \BibitemOpen
  \bibfield  {author} {\bibinfo {author} {\bibfnamefont {Y.}~\bibnamefont
  {Kohama}}, \bibinfo {author} {\bibfnamefont {K.}~\bibnamefont {Mochidzuki}},
  \bibinfo {author} {\bibfnamefont {T.}~\bibnamefont {Terashima}}, \bibinfo
  {author} {\bibfnamefont {A.}~\bibnamefont {Miyata}}, \bibinfo {author}
  {\bibfnamefont {A.}~\bibnamefont {DeMuer}}, \bibinfo {author} {\bibfnamefont
  {T.}~\bibnamefont {Klein}}, \bibinfo {author} {\bibfnamefont
  {C.}~\bibnamefont {Marcenat}}, \bibinfo {author} {\bibfnamefont {Z.~L.}\
  \bibnamefont {Dun}}, \bibinfo {author} {\bibfnamefont {H.}~\bibnamefont
  {Zhou}}, \bibinfo {author} {\bibfnamefont {G.}~\bibnamefont {Li}}, \bibinfo
  {author} {\bibfnamefont {L.}~\bibnamefont {Balicas}}, \bibinfo {author}
  {\bibfnamefont {N.}~\bibnamefont {Abe}}, \bibinfo {author} {\bibfnamefont
  {Y.~H.}\ \bibnamefont {Matsuda}}, \bibinfo {author} {\bibfnamefont
  {S.}~\bibnamefont {Takeyama}}, \bibinfo {author} {\bibfnamefont
  {A.}~\bibnamefont {Matsuo}}, \ and\ \bibinfo {author} {\bibfnamefont
  {K.}~\bibnamefont {Kindo}},\ }\bibfield  {title} {\enquote {\bibinfo {title}
  {Entropy of the quantum soliton lattice and multiple magnetization steps in
  {${\mathrm{BiCu}}_{2}{\mathrm{PO}}_{6}$}},}\ }\href {\doibase
  10.1103/PhysRevB.90.060408} {\bibfield  {journal} {\bibinfo  {journal} {Phys.
  Rev. B}\ }\textbf {\bibinfo {volume} {90}},\ \bibinfo {pages} {060408(R)}
  (\bibinfo {year} {2014})}\BibitemShut {NoStop}%
\bibitem [{\citenamefont {Koteswararao}\ \emph {et~al.}(2007)\citenamefont
  {Koteswararao}, \citenamefont {Salunke}, \citenamefont {Mahajan},
  \citenamefont {Dasgupta},\ and\ \citenamefont {Bobroff}}]{Koteswararao2007}%
  \BibitemOpen
  \bibfield  {author} {\bibinfo {author} {\bibfnamefont {B.}~\bibnamefont
  {Koteswararao}}, \bibinfo {author} {\bibfnamefont {S.}~\bibnamefont
  {Salunke}}, \bibinfo {author} {\bibfnamefont {A.~V.}\ \bibnamefont
  {Mahajan}}, \bibinfo {author} {\bibfnamefont {I.}~\bibnamefont {Dasgupta}}, \
  and\ \bibinfo {author} {\bibfnamefont {J.}~\bibnamefont {Bobroff}},\
  }\bibfield  {title} {\enquote {\bibinfo {title} {Spin-gap behavior in the
  two-leg spin-ladder {$\text{Bi}{\text{Cu}}_{2}\text{P}{\text{O}}_{6}$}},}\
  }\href {\doibase 10.1103/PhysRevB.76.052402} {\bibfield  {journal} {\bibinfo
  {journal} {Phys. Rev. B}\ }\textbf {\bibinfo {volume} {76}},\ \bibinfo
  {pages} {052402} (\bibinfo {year} {2007})}\BibitemShut {NoStop}%
\bibitem [{\citenamefont {Mentr\'e}\ \emph {et~al.}(2009)\citenamefont
  {Mentr\'e}, \citenamefont {Janod}, \citenamefont {Rabu}, \citenamefont
  {Hennion}, \citenamefont {Leclercq-Hugeux}, \citenamefont {Kang},
  \citenamefont {Lee}, \citenamefont {Whangbo},\ and\ \citenamefont
  {Petit}}]{Mentre2009}%
  \BibitemOpen
  \bibfield  {author} {\bibinfo {author} {\bibfnamefont {O.}~\bibnamefont
  {Mentr\'e}}, \bibinfo {author} {\bibfnamefont {E.}~\bibnamefont {Janod}},
  \bibinfo {author} {\bibfnamefont {P.}~\bibnamefont {Rabu}}, \bibinfo {author}
  {\bibfnamefont {M.}~\bibnamefont {Hennion}}, \bibinfo {author} {\bibfnamefont
  {F.}~\bibnamefont {Leclercq-Hugeux}}, \bibinfo {author} {\bibfnamefont
  {J.}~\bibnamefont {Kang}}, \bibinfo {author} {\bibfnamefont {C.}~\bibnamefont
  {Lee}}, \bibinfo {author} {\bibfnamefont {M.-H.}\ \bibnamefont {Whangbo}}, \
  and\ \bibinfo {author} {\bibfnamefont {S.}~\bibnamefont {Petit}},\ }\bibfield
   {title} {\enquote {\bibinfo {title} {Incommensurate spin correlation driven
  by frustration in {${\text{BiCu}}_{2}{\text{PO}}_{6}$}},}\ }\href {\doibase
  10.1103/PhysRevB.80.180413} {\bibfield  {journal} {\bibinfo  {journal} {Phys.
  Rev. B}\ }\textbf {\bibinfo {volume} {80}},\ \bibinfo {pages} {180413(R)}
  (\bibinfo {year} {2009})}\BibitemShut {NoStop}%
\bibitem [{\citenamefont {Plumb}\ \emph {et~al.}(2013)\citenamefont {Plumb},
  \citenamefont {Yamani}, \citenamefont {Matsuda}, \citenamefont {Shu},
  \citenamefont {Koteswararao}, \citenamefont {Chou},\ and\ \citenamefont
  {Kim}}]{Plumb2013}%
  \BibitemOpen
  \bibfield  {author} {\bibinfo {author} {\bibfnamefont {K.~W.}\ \bibnamefont
  {Plumb}}, \bibinfo {author} {\bibfnamefont {Z.}~\bibnamefont {Yamani}},
  \bibinfo {author} {\bibfnamefont {M.}~\bibnamefont {Matsuda}}, \bibinfo
  {author} {\bibfnamefont {G.~J.}\ \bibnamefont {Shu}}, \bibinfo {author}
  {\bibfnamefont {B.}~\bibnamefont {Koteswararao}}, \bibinfo {author}
  {\bibfnamefont {F.~C.}\ \bibnamefont {Chou}}, \ and\ \bibinfo {author}
  {\bibfnamefont {Y.-J.}\ \bibnamefont {Kim}},\ }\bibfield  {title} {\enquote
  {\bibinfo {title} {Incommensurate dynamic correlations in the
  quasi-two-dimensional spin liquid {BiCu${}_{2}$PO${}_{6}$}},}\ }\href
  {\doibase 10.1103/PhysRevB.88.024402} {\bibfield  {journal} {\bibinfo
  {journal} {Phys. Rev. B}\ }\textbf {\bibinfo {volume} {88}},\ \bibinfo
  {pages} {024402} (\bibinfo {year} {2013})}\BibitemShut {NoStop}%
\bibitem [{\citenamefont {Splinter}\ \emph {et~al.}(2016)\citenamefont
  {Splinter}, \citenamefont {Drescher}, \citenamefont {Krull},\ and\
  \citenamefont {Uhrig}}]{Splinter2016}%
  \BibitemOpen
  \bibfield  {author} {\bibinfo {author} {\bibfnamefont {L.}~\bibnamefont
  {Splinter}}, \bibinfo {author} {\bibfnamefont {N.~A.}\ \bibnamefont
  {Drescher}}, \bibinfo {author} {\bibfnamefont {H.}~\bibnamefont {Krull}}, \
  and\ \bibinfo {author} {\bibfnamefont {G.~S.}\ \bibnamefont {Uhrig}},\
  }\bibfield  {title} {\enquote {\bibinfo {title} {Minimal model for the
  frustrated spin ladder system {${\mathrm{BiCu}}_{2}{\mathrm{PO}}_{6}$}},}\
  }\href {\doibase 10.1103/PhysRevB.94.155115} {\bibfield  {journal} {\bibinfo
  {journal} {Phys. Rev. B}\ }\textbf {\bibinfo {volume} {94}},\ \bibinfo
  {pages} {155115} (\bibinfo {year} {2016})}\BibitemShut {NoStop}%
\bibitem [{\citenamefont {Dzyaloshinsky}(1958)}]{Dzyaloshinsky1958}%
  \BibitemOpen
  \bibfield  {author} {\bibinfo {author} {\bibfnamefont {I.}~\bibnamefont
  {Dzyaloshinsky}},\ }\bibfield  {title} {\enquote {\bibinfo {title} {A
  thermodynamic theory of “weak” ferromagnetism of antiferromagnetics},}\
  }\href {\doibase 10.1016/0022-3697(58)90076-3} {\bibfield  {journal}
  {\bibinfo  {journal} {J. Phys. Chem. Solids}\ }\textbf {\bibinfo {volume}
  {4}},\ \bibinfo {pages} {241--255} (\bibinfo {year} {1958})}\BibitemShut
  {NoStop}%
\bibitem [{\citenamefont {Moriya}(1960)}]{Moriya1960_1}%
  \BibitemOpen
  \bibfield  {author} {\bibinfo {author} {\bibfnamefont {T.}~\bibnamefont
  {Moriya}},\ }\bibfield  {title} {\enquote {\bibinfo {title} {New mechanism of
  anisotropic superexchange interaction},}\ }\href {\doibase
  10.1103/PhysRevLett.4.228} {\bibfield  {journal} {\bibinfo  {journal} {Phys.
  Rev. Lett.}\ }\textbf {\bibinfo {volume} {4}},\ \bibinfo {pages} {228--230}
  (\bibinfo {year} {1960})}\BibitemShut {NoStop}%
\bibitem [{\citenamefont {Sudan}\ \emph {et~al.}(2009)\citenamefont {Sudan},
  \citenamefont {L\"uscher},\ and\ \citenamefont {L\"auchli}}]{Sudan2009}%
  \BibitemOpen
  \bibfield  {author} {\bibinfo {author} {\bibfnamefont {J.}~\bibnamefont
  {Sudan}}, \bibinfo {author} {\bibfnamefont {A.}~\bibnamefont {L\"uscher}}, \
  and\ \bibinfo {author} {\bibfnamefont {A.~M.}\ \bibnamefont {L\"auchli}},\
  }\bibfield  {title} {\enquote {\bibinfo {title} {Emergent multipolar spin
  correlations in a fluctuating spiral: {T}he frustrated ferromagnetic
  spin-{$\frac{1}{2}$} {H}eisenberg chain in a magnetic field},}\ }\href
  {\doibase 10.1103/PhysRevB.80.140402} {\bibfield  {journal} {\bibinfo
  {journal} {Phys. Rev. B}\ }\textbf {\bibinfo {volume} {80}},\ \bibinfo
  {pages} {140402(R)} (\bibinfo {year} {2009})}\BibitemShut {NoStop}%
\bibitem [{\citenamefont {Jeon}\ \emph {et~al.}(2014)\citenamefont {Jeon},
  \citenamefont {Koteswararao}, \citenamefont {Hoon~Kim}, \citenamefont {Kim},
  \citenamefont {Shu},\ and\ \citenamefont {Chou}}]{Jeon2014}%
  \BibitemOpen
  \bibfield  {author} {\bibinfo {author} {\bibfnamefont {B.-G.}\ \bibnamefont
  {Jeon}}, \bibinfo {author} {\bibfnamefont {B.}~\bibnamefont {Koteswararao}},
  \bibinfo {author} {\bibfnamefont {K.}~\bibnamefont {Hoon~Kim}}, \bibinfo
  {author} {\bibfnamefont {J.~W.}\ \bibnamefont {Kim}}, \bibinfo {author}
  {\bibfnamefont {G.~J.}\ \bibnamefont {Shu}}, \ and\ \bibinfo {author}
  {\bibfnamefont {F.~C.}\ \bibnamefont {Chou}},\ }\href
  {https://legacywww.magnet.fsu.edu/mediacenter/publications/reports/2014annualreport/2014-NHMFL-Report223.pdf}
  {\emph {\bibinfo {title} {Multiferroicity in a {F}rustrated {S}pin {L}adder
  {BiCu$_2$PO$_6$} {A}t {H}igh {M}agnetic {F}ield}}},\ \bibinfo {type}
  {Research Report}\ \bibinfo {number} {223}\ (\bibinfo  {institution}
  {National High Magnetic Field Laboratory},\ \bibinfo {address} {Tallahassee,
  FL, USA},\ \bibinfo {year} {2014})\BibitemShut {NoStop}%
\bibitem [{\citenamefont {Kaplan}(1983)}]{Kaplan1983}%
  \BibitemOpen
  \bibfield  {author} {\bibinfo {author} {\bibfnamefont {T.~A.}\ \bibnamefont
  {Kaplan}},\ }\bibfield  {title} {\enquote {\bibinfo {title} {Single-{B}and
  {H}ubbard {M}odel with {S}pin-{O}rbit {C}oupling},}\ }\href {\doibase
  10.1007/BF01301591} {\bibfield  {journal} {\bibinfo  {journal} {Z. Phys. B
  Con. Mat.}\ }\textbf {\bibinfo {volume} {49}},\ \bibinfo {pages} {313--317}
  (\bibinfo {year} {1983})}\BibitemShut {NoStop}%
\bibitem [{\citenamefont {Shekhtman}\ \emph {et~al.}(1992)\citenamefont
  {Shekhtman}, \citenamefont {Entin-Wohlman},\ and\ \citenamefont
  {Aharony}}]{Shektman1992}%
  \BibitemOpen
  \bibfield  {author} {\bibinfo {author} {\bibfnamefont {L.}~\bibnamefont
  {Shekhtman}}, \bibinfo {author} {\bibfnamefont {O.}~\bibnamefont
  {Entin-Wohlman}}, \ and\ \bibinfo {author} {\bibfnamefont {A.}~\bibnamefont
  {Aharony}},\ }\bibfield  {title} {\enquote {\bibinfo {title} {Moriya's
  {A}nisotropic {S}uperexchange {I}nteraction, {F}rustration, and
  {D}zyaloshinsky's {W}eak {F}erromagnetism},}\ }\href {\doibase
  10.1103/PhysRevLett.69.836} {\bibfield  {journal} {\bibinfo  {journal} {Phys.
  Rev. Lett.}\ }\textbf {\bibinfo {volume} {69}},\ \bibinfo {pages} {836--839}
  (\bibinfo {year} {1992})}\BibitemShut {NoStop}%
\bibitem [{\citenamefont {Veillette}\ \emph {et~al.}(2005)\citenamefont
  {Veillette}, \citenamefont {Chalker},\ and\ \citenamefont
  {Coldea}}]{Veillette2005}%
  \BibitemOpen
  \bibfield  {author} {\bibinfo {author} {\bibfnamefont {M.~Y.}\ \bibnamefont
  {Veillette}}, \bibinfo {author} {\bibfnamefont {J.~T.}\ \bibnamefont
  {Chalker}}, \ and\ \bibinfo {author} {\bibfnamefont {R.}~\bibnamefont
  {Coldea}},\ }\bibfield  {title} {\enquote {\bibinfo {title} {Ground states of
  a frustrated spin-{$\frac{1}{2}$} antiferromagnet:
  {$\mathrm{Cs}_{2}\mathrm{Cu}\mathrm{Cl}_{4}$} in a magnetic field},}\ }\href
  {\doibase 10.1103/PhysRevB.71.214426} {\bibfield  {journal} {\bibinfo
  {journal} {Phys. Rev. B}\ }\textbf {\bibinfo {volume} {71}},\ \bibinfo
  {pages} {214426} (\bibinfo {year} {2005})}\BibitemShut {NoStop}%
\bibitem [{\citenamefont {Blinc}(1981)}]{Blinc1981}%
  \BibitemOpen
  \bibfield  {author} {\bibinfo {author} {\bibfnamefont {R.}~\bibnamefont
  {Blinc}},\ }\bibfield  {title} {\enquote {\bibinfo {title} {Magnetic
  resonance and relaxation in structurally incommensurate systems},}\ }\href
  {\doibase 10.1016/0370-1573(81)90108-3} {\bibfield  {journal} {\bibinfo
  {journal} {Phys. Rep.}\ }\textbf {\bibinfo {volume} {79}},\ \bibinfo {pages}
  {331--398} (\bibinfo {year} {1981})}\BibitemShut {NoStop}%
\bibitem [{\citenamefont {Choi}\ \emph {et~al.}(2013)\citenamefont {Choi},
  \citenamefont {Hwang}, \citenamefont {Lemmens}, \citenamefont {Wulferding},
  \citenamefont {Shu},\ and\ \citenamefont {Chou}}]{Choi2013}%
  \BibitemOpen
  \bibfield  {author} {\bibinfo {author} {\bibfnamefont {K.-Y.}\ \bibnamefont
  {Choi}}, \bibinfo {author} {\bibfnamefont {J.~W.}\ \bibnamefont {Hwang}},
  \bibinfo {author} {\bibfnamefont {P.}~\bibnamefont {Lemmens}}, \bibinfo
  {author} {\bibfnamefont {D.}~\bibnamefont {Wulferding}}, \bibinfo {author}
  {\bibfnamefont {G.~J.}\ \bibnamefont {Shu}}, \ and\ \bibinfo {author}
  {\bibfnamefont {F.~C.}\ \bibnamefont {Chou}},\ }\bibfield  {title} {\enquote
  {\bibinfo {title} {Evidence for {D}imer {C}rystal {M}elting in the
  {F}rustrated {S}pin-{L}adder {S}ystem
  {${\mathrm{BiCu}}_{2}{\mathrm{PO}}_{6}$}},}\ }\href {\doibase
  10.1103/PhysRevLett.110.117204} {\bibfield  {journal} {\bibinfo  {journal}
  {Phys. Rev. Lett.}\ }\textbf {\bibinfo {volume} {110}},\ \bibinfo {pages}
  {117204} (\bibinfo {year} {2013})}\BibitemShut {NoStop}%
\bibitem [{\citenamefont {Cheong}\ and\ \citenamefont
  {Mostovoy}(2007)}]{Cheong2007}%
  \BibitemOpen
  \bibfield  {author} {\bibinfo {author} {\bibfnamefont {S.-W.}\ \bibnamefont
  {Cheong}}\ and\ \bibinfo {author} {\bibfnamefont {M.}~\bibnamefont
  {Mostovoy}},\ }\bibfield  {title} {\enquote {\bibinfo {title} {Multiferroics:
  a magnetic twist for ferroelectricity},}\ }\href {\doibase 10.1038/nmat1804}
  {\bibfield  {journal} {\bibinfo  {journal} {Nat. Mater.}\ }\textbf {\bibinfo
  {volume} {6}},\ \bibinfo {pages} {13--20} (\bibinfo {year}
  {2007})}\BibitemShut {NoStop}%
\bibitem [{\citenamefont {Bar'yakhtar}\ \emph {et~al.}(1983)\citenamefont
  {Bar'yakhtar}, \citenamefont {L'vov},\ and\ \citenamefont
  {Yablonskii}}]{Baryakhtar1983}%
  \BibitemOpen
  \bibfield  {author} {\bibinfo {author} {\bibfnamefont {V.~G.}\ \bibnamefont
  {Bar'yakhtar}}, \bibinfo {author} {\bibfnamefont {V.~A.}\ \bibnamefont
  {L'vov}}, \ and\ \bibinfo {author} {\bibfnamefont {D.~A.}\ \bibnamefont
  {Yablonskii}},\ }\bibfield  {title} {\enquote {\bibinfo {title}
  {Inhomogeneous magnetoelectric effect},}\ }\href
  {http://www.jetpletters.ac.ru/ps/1499/article_22895.shtml} {\bibfield
  {journal} {\bibinfo  {journal} {JETP Lett.}\ }\textbf {\bibinfo {volume}
  {37}},\ \bibinfo {pages} {673--675} (\bibinfo {year} {1983})}\BibitemShut
  {NoStop}%
\bibitem [{\citenamefont {Furukawa}\ \emph {et~al.}(2008)\citenamefont
  {Furukawa}, \citenamefont {Sato}, \citenamefont {Saiga},\ and\ \citenamefont
  {Onoda}}]{Furukawa2008}%
  \BibitemOpen
  \bibfield  {author} {\bibinfo {author} {\bibfnamefont {S.}~\bibnamefont
  {Furukawa}}, \bibinfo {author} {\bibfnamefont {M.}~\bibnamefont {Sato}},
  \bibinfo {author} {\bibfnamefont {Y.}~\bibnamefont {Saiga}}, \ and\ \bibinfo
  {author} {\bibfnamefont {S.}~\bibnamefont {Onoda}},\ }\bibfield  {title}
  {\enquote {\bibinfo {title} {Quantum {F}luctuations of {C}hirality in
  {O}ne-{D}imensional {S}pin-1/2 {M}ultiferroics: {G}apless {D}ielectric
  {R}esponse from {P}hasons and {C}hiral {S}olitons},}\ }\href {\doibase
  10.1143/JPSJ.77.123712} {\bibfield  {journal} {\bibinfo  {journal} {J. Phys.
  Soc. Jpn.}\ }\textbf {\bibinfo {volume} {77}},\ \bibinfo {pages} {123712}
  (\bibinfo {year} {2008})}\BibitemShut {NoStop}%
\bibitem [{\citenamefont {Khomskii}\ \emph {et~al.}(1996)\citenamefont
  {Khomskii}, \citenamefont {Geertsma},\ and\ \citenamefont
  {Mostovoy}}]{Khomskii1996}%
  \BibitemOpen
  \bibfield  {author} {\bibinfo {author} {\bibfnamefont {D.}~\bibnamefont
  {Khomskii}}, \bibinfo {author} {\bibfnamefont {W.}~\bibnamefont {Geertsma}},
  \ and\ \bibinfo {author} {\bibfnamefont {M.}~\bibnamefont {Mostovoy}},\
  }\bibfield  {title} {\enquote {\bibinfo {title} {Elementary excitations,
  exchange interaction and spin-{P}eierls transition in
  {$\mathrm{CuGeO}_3$}},}\ }\href {\doibase 10.1007/BF02548136} {\bibfield
  {journal} {\bibinfo  {journal} {Czech. J. Phys.}\ }\textbf {\bibinfo {volume}
  {46}},\ \bibinfo {pages} {3239--3246} (\bibinfo {year} {1996})}\BibitemShut
  {NoStop}%
\bibitem [{\citenamefont {Zang}\ \emph {et~al.}(1997)\citenamefont {Zang},
  \citenamefont {Chakravarty},\ and\ \citenamefont {Bishop}}]{Zang1997}%
  \BibitemOpen
  \bibfield  {author} {\bibinfo {author} {\bibfnamefont {J.}~\bibnamefont
  {Zang}}, \bibinfo {author} {\bibfnamefont {S.}~\bibnamefont {Chakravarty}}, \
  and\ \bibinfo {author} {\bibfnamefont {A.~R.}\ \bibnamefont {Bishop}},\
  }\bibfield  {title} {\enquote {\bibinfo {title} {Interchain coupling effects
  and solitons in {${\mathrm{CuGeO}}_{3}$}},}\ }\href {\doibase
  10.1103/PhysRevB.55.R14705} {\bibfield  {journal} {\bibinfo  {journal} {Phys.
  Rev. B}\ }\textbf {\bibinfo {volume} {55}},\ \bibinfo {pages}
  {R14705--R14708} (\bibinfo {year} {1997})}\BibitemShut {NoStop}%
\bibitem [{\citenamefont {Dobry}\ and\ \citenamefont
  {Riera}(1997)}]{Dobry1997}%
  \BibitemOpen
  \bibfield  {author} {\bibinfo {author} {\bibfnamefont {A.}~\bibnamefont
  {Dobry}}\ and\ \bibinfo {author} {\bibfnamefont {J.~A.}\ \bibnamefont
  {Riera}},\ }\bibfield  {title} {\enquote {\bibinfo {title} {Soliton width in
  the incommensurate phase of spin-{P}eierls systems},}\ }\href {\doibase
  10.1103/PhysRevB.56.R2912} {\bibfield  {journal} {\bibinfo  {journal} {Phys.
  Rev. B}\ }\textbf {\bibinfo {volume} {56}},\ \bibinfo {pages} {R2912--R2915}
  (\bibinfo {year} {1997})}\BibitemShut {NoStop}%
\bibitem [{\citenamefont {S\o{}rensen}\ \emph {et~al.}(1998)\citenamefont
  {S\o{}rensen}, \citenamefont {Affleck}, \citenamefont {Augier},\ and\
  \citenamefont {Poilblanc}}]{Sorensen1998}%
  \BibitemOpen
  \bibfield  {author} {\bibinfo {author} {\bibfnamefont {E.}~\bibnamefont
  {S\o{}rensen}}, \bibinfo {author} {\bibfnamefont {I.}~\bibnamefont
  {Affleck}}, \bibinfo {author} {\bibfnamefont {D.}~\bibnamefont {Augier}}, \
  and\ \bibinfo {author} {\bibfnamefont {D.}~\bibnamefont {Poilblanc}},\
  }\bibfield  {title} {\enquote {\bibinfo {title} {Soliton approach to
  spin-{P}eierls antiferromagnets: Large-scale numerical results},}\ }\href
  {\doibase 10.1103/PhysRevB.58.R14701} {\bibfield  {journal} {\bibinfo
  {journal} {Phys. Rev. B}\ }\textbf {\bibinfo {volume} {58}},\ \bibinfo
  {pages} {R14701--R14704} (\bibinfo {year} {1998})}\BibitemShut {NoStop}%
\bibitem [{\citenamefont {Uhrig}\ \emph {et~al.}(1999)\citenamefont {Uhrig},
  \citenamefont {Sch\"onfeld}, \citenamefont {Boucher},\ and\ \citenamefont
  {Horvati\'{c}}}]{Uhrig1999}%
  \BibitemOpen
  \bibfield  {author} {\bibinfo {author} {\bibfnamefont {G.~S.}\ \bibnamefont
  {Uhrig}}, \bibinfo {author} {\bibfnamefont {F.}~\bibnamefont {Sch\"onfeld}},
  \bibinfo {author} {\bibfnamefont {J.-P.}\ \bibnamefont {Boucher}}, \ and\
  \bibinfo {author} {\bibfnamefont {M.}~\bibnamefont {Horvati\'{c}}},\
  }\bibfield  {title} {\enquote {\bibinfo {title} {Soliton lattices in the
  incommensurate spin-{P}eierls phase: Local distortions and magnetizations},}\
  }\href {\doibase 10.1103/PhysRevB.60.9468} {\bibfield  {journal} {\bibinfo
  {journal} {Phys. Rev. B}\ }\textbf {\bibinfo {volume} {60}},\ \bibinfo
  {pages} {9468--9476} (\bibinfo {year} {1999})}\BibitemShut {NoStop}%
\bibitem [{\citenamefont {Onoda}\ and\ \citenamefont
  {Nagaosa}(2007)}]{Onoda2007}%
  \BibitemOpen
  \bibfield  {author} {\bibinfo {author} {\bibfnamefont {Shigeki}\ \bibnamefont
  {Onoda}}\ and\ \bibinfo {author} {\bibfnamefont {Naoto}\ \bibnamefont
  {Nagaosa}},\ }\bibfield  {title} {\enquote {\bibinfo {title} {Chiral {S}pin
  {P}airing in {H}elical {M}agnets},}\ }\href {\doibase
  10.1103/PhysRevLett.99.027206} {\bibfield  {journal} {\bibinfo  {journal}
  {Phys. Rev. Lett.}\ }\textbf {\bibinfo {volume} {99}},\ \bibinfo {pages}
  {027206} (\bibinfo {year} {2007})}\BibitemShut {NoStop}%
\bibitem [{\citenamefont {Sergienko}\ and\ \citenamefont
  {Dagotto}(2006)}]{Sergienko2006_MC}%
  \BibitemOpen
  \bibfield  {author} {\bibinfo {author} {\bibfnamefont {I.~A.}\ \bibnamefont
  {Sergienko}}\ and\ \bibinfo {author} {\bibfnamefont {E.}~\bibnamefont
  {Dagotto}},\ }\bibfield  {title} {\enquote {\bibinfo {title} {Role of the
  {D}zyaloshinskii-{M}oriya interaction in multiferroic perovskites},}\ }\href
  {\doibase 10.1103/PhysRevB.73.094434} {\bibfield  {journal} {\bibinfo
  {journal} {Phys. Rev. B}\ }\textbf {\bibinfo {volume} {73}},\ \bibinfo
  {pages} {094434} (\bibinfo {year} {2006})}\BibitemShut {NoStop}%
\bibitem [{\citenamefont {Pikulski}(2017)}]{DissPikulski}%
  \BibitemOpen
  \bibfield  {author} {\bibinfo {author} {\bibfnamefont {M.}~\bibnamefont
  {Pikulski}},\ }\emph {\bibinfo {title} {Field-induced chirality in a
  frustrated quantum spin ladder}},\ \href {\doibase 10.3929/ethz-b-000204120}
  {Ph.D. thesis},\ \bibinfo  {school} {ETH Zürich}, \bibinfo {address}
  {Z\"urich, Switzerland} (\bibinfo {year} {2017})\BibitemShut {NoStop}%
\bibitem [{\citenamefont {White}(1992)}]{White1992}%
  \BibitemOpen
  \bibfield  {author} {\bibinfo {author} {\bibfnamefont {S.~R.}\ \bibnamefont
  {White}},\ }\bibfield  {title} {\enquote {\bibinfo {title} {Density {M}atrix
  {F}ormulation for {Q}uantum {R}enormalization {G}roups},}\ }\href {\doibase
  10.1103/PhysRevLett.69.2863} {\bibfield  {journal} {\bibinfo  {journal}
  {Phys. Rev. Lett.}\ }\textbf {\bibinfo {volume} {69}},\ \bibinfo {pages}
  {2863--2866} (\bibinfo {year} {1992})}\BibitemShut {NoStop}%
\bibitem [{\citenamefont {Dolfi}\ \emph {et~al.}(2015)\citenamefont {Dolfi},
  \citenamefont {Bauer}, \citenamefont {Keller},\ and\ \citenamefont
  {Troyer}}]{Dolfi2015}%
  \BibitemOpen
  \bibfield  {author} {\bibinfo {author} {\bibfnamefont {M.}~\bibnamefont
  {Dolfi}}, \bibinfo {author} {\bibfnamefont {B.}~\bibnamefont {Bauer}},
  \bibinfo {author} {\bibfnamefont {S.}~\bibnamefont {Keller}}, \ and\ \bibinfo
  {author} {\bibfnamefont {M.}~\bibnamefont {Troyer}},\ }\bibfield  {title}
  {\enquote {\bibinfo {title} {Pair correlations in doped {H}ubbard ladders},}\
  }\href {\doibase 10.1103/PhysRevB.92.195139} {\bibfield  {journal} {\bibinfo
  {journal} {Phys. Rev. B}\ }\textbf {\bibinfo {volume} {92}},\ \bibinfo
  {pages} {195139} (\bibinfo {year} {2015})}\BibitemShut {NoStop}%
\bibitem [{\citenamefont {Dolfi}\ \emph {et~al.}(2014)\citenamefont {Dolfi},
  \citenamefont {Bauer}, \citenamefont {Keller}, \citenamefont {Kosenkov},
  \citenamefont {Ewart}, \citenamefont {Kantian}, \citenamefont {Giamarchi},\
  and\ \citenamefont {Troyer}}]{Dolfi2014}%
  \BibitemOpen
  \bibfield  {author} {\bibinfo {author} {\bibfnamefont {M.}~\bibnamefont
  {Dolfi}}, \bibinfo {author} {\bibfnamefont {B.}~\bibnamefont {Bauer}},
  \bibinfo {author} {\bibfnamefont {S.}~\bibnamefont {Keller}}, \bibinfo
  {author} {\bibfnamefont {A.}~\bibnamefont {Kosenkov}}, \bibinfo {author}
  {\bibfnamefont {T.}~\bibnamefont {Ewart}}, \bibinfo {author} {\bibfnamefont
  {A.}~\bibnamefont {Kantian}}, \bibinfo {author} {\bibfnamefont {Th.}\
  \bibnamefont {Giamarchi}}, \ and\ \bibinfo {author} {\bibfnamefont
  {M.}~\bibnamefont {Troyer}},\ }\bibfield  {title} {\enquote {\bibinfo {title}
  {Matrix product state applications for the {ALPS} project},}\ }\href
  {\doibase 10.1016/j.cpc.2014.08.019} {\bibfield  {journal} {\bibinfo
  {journal} {Comput. Phys. Commun.}\ }\textbf {\bibinfo {volume} {185}},\
  \bibinfo {pages} {3430--3440} (\bibinfo {year} {2014})}\BibitemShut {NoStop}%
\bibitem [{\citenamefont {Oshikawa}\ and\ \citenamefont
  {Affleck}(1997)}]{Oshikawa1997}%
  \BibitemOpen
  \bibfield  {author} {\bibinfo {author} {\bibfnamefont {M.}~\bibnamefont
  {Oshikawa}}\ and\ \bibinfo {author} {\bibfnamefont {I.}~\bibnamefont
  {Affleck}},\ }\bibfield  {title} {\enquote {\bibinfo {title} {Field-{I}nduced
  {G}ap in {$S=1/2$} {A}ntiferromagnetic {C}hains},}\ }\href {\doibase
  10.1103/PhysRevLett.79.2883} {\bibfield  {journal} {\bibinfo  {journal}
  {Phys. Rev. Lett.}\ }\textbf {\bibinfo {volume} {79}},\ \bibinfo {pages}
  {2883--2886} (\bibinfo {year} {1997})}\BibitemShut {NoStop}%
\bibitem [{\citenamefont {C\'epas}\ \emph {et~al.}(2001)\citenamefont
  {C\'epas}, \citenamefont {Kakurai}, \citenamefont {Regnault}, \citenamefont
  {Ziman}, \citenamefont {Boucher}, \citenamefont {Aso}, \citenamefont {Nishi},
  \citenamefont {Kageyama},\ and\ \citenamefont {Ueda}}]{Cepas2001}%
  \BibitemOpen
  \bibfield  {author} {\bibinfo {author} {\bibfnamefont {O.}~\bibnamefont
  {C\'epas}}, \bibinfo {author} {\bibfnamefont {K.}~\bibnamefont {Kakurai}},
  \bibinfo {author} {\bibfnamefont {L.~P.}\ \bibnamefont {Regnault}}, \bibinfo
  {author} {\bibfnamefont {T.}~\bibnamefont {Ziman}}, \bibinfo {author}
  {\bibfnamefont {J.~P.}\ \bibnamefont {Boucher}}, \bibinfo {author}
  {\bibfnamefont {N.}~\bibnamefont {Aso}}, \bibinfo {author} {\bibfnamefont
  {M.}~\bibnamefont {Nishi}}, \bibinfo {author} {\bibfnamefont
  {H.}~\bibnamefont {Kageyama}}, \ and\ \bibinfo {author} {\bibfnamefont
  {Y.}~\bibnamefont {Ueda}},\ }\bibfield  {title} {\enquote {\bibinfo {title}
  {Dzyaloshinski-{M}oriya {I}nteraction in the {2D} {S}pin {G}ap {S}ystem
  {${\mathrm{SrCu}}_{2}({\mathrm{BO}}_{3}{)}_{2}$}},}\ }\href {\doibase
  10.1103/PhysRevLett.87.167205} {\bibfield  {journal} {\bibinfo  {journal}
  {Phys. Rev. Lett.}\ }\textbf {\bibinfo {volume} {87}},\ \bibinfo {pages}
  {167205} (\bibinfo {year} {2001})}\BibitemShut {NoStop}%
\bibitem [{\citenamefont {Chernyshev}(2005)}]{Chernyshev2005}%
  \BibitemOpen
  \bibfield  {author} {\bibinfo {author} {\bibfnamefont {A.~L.}\ \bibnamefont
  {Chernyshev}},\ }\bibfield  {title} {\enquote {\bibinfo {title} {Effects of
  an external magnetic field on the gaps and quantum corrections in an ordered
  {H}eisenberg antiferromagnet with {D}zyaloshinskii-{M}oriya anisotropy},}\
  }\href {\doibase 10.1103/PhysRevB.72.174414} {\bibfield  {journal} {\bibinfo
  {journal} {Phys. Rev. B}\ }\textbf {\bibinfo {volume} {72}},\ \bibinfo
  {pages} {174414} (\bibinfo {year} {2005})}\BibitemShut {NoStop}%
\bibitem [{\citenamefont {Fouet}\ \emph {et~al.}(2006)\citenamefont {Fouet},
  \citenamefont {Mila}, \citenamefont {Clarke}, \citenamefont {Youk},
  \citenamefont {Tchernyshyov}, \citenamefont {Fendley},\ and\ \citenamefont
  {Noack}}]{Fouet2006}%
  \BibitemOpen
  \bibfield  {author} {\bibinfo {author} {\bibfnamefont {J.-B.}\ \bibnamefont
  {Fouet}}, \bibinfo {author} {\bibfnamefont {F.}~\bibnamefont {Mila}},
  \bibinfo {author} {\bibfnamefont {D.}~\bibnamefont {Clarke}}, \bibinfo
  {author} {\bibfnamefont {H.}~\bibnamefont {Youk}}, \bibinfo {author}
  {\bibfnamefont {O.}~\bibnamefont {Tchernyshyov}}, \bibinfo {author}
  {\bibfnamefont {P.}~\bibnamefont {Fendley}}, \ and\ \bibinfo {author}
  {\bibfnamefont {R.~M.}\ \bibnamefont {Noack}},\ }\bibfield  {title} {\enquote
  {\bibinfo {title} {Condensation of magnons and spinons in a frustrated
  ladder},}\ }\href {\doibase 10.1103/PhysRevB.73.214405} {\bibfield  {journal}
  {\bibinfo  {journal} {Phys. Rev. B}\ }\textbf {\bibinfo {volume} {73}},\
  \bibinfo {pages} {214405} (\bibinfo {year} {2006})}\BibitemShut {NoStop}%
\bibitem [{\citenamefont {Hao}\ \emph {et~al.}(2011)\citenamefont {Hao},
  \citenamefont {Wan}, \citenamefont {Rousochatzakis}, \citenamefont
  {Wildeboer}, \citenamefont {Seidel}, \citenamefont {Mila},\ and\
  \citenamefont {Tchernyshyov}}]{Hao2011}%
  \BibitemOpen
  \bibfield  {author} {\bibinfo {author} {\bibfnamefont {Z.}~\bibnamefont
  {Hao}}, \bibinfo {author} {\bibfnamefont {Y.}~\bibnamefont {Wan}}, \bibinfo
  {author} {\bibfnamefont {I.}~\bibnamefont {Rousochatzakis}}, \bibinfo
  {author} {\bibfnamefont {J.}~\bibnamefont {Wildeboer}}, \bibinfo {author}
  {\bibfnamefont {A.}~\bibnamefont {Seidel}}, \bibinfo {author} {\bibfnamefont
  {F.}~\bibnamefont {Mila}}, \ and\ \bibinfo {author} {\bibfnamefont
  {O.}~\bibnamefont {Tchernyshyov}},\ }\bibfield  {title} {\enquote {\bibinfo
  {title} {Destruction of valence-bond order in a {$S=1/2$} sawtooth chain with
  a {D}zyaloshinskii-{M}oriya term},}\ }\href {\doibase
  10.1103/PhysRevB.84.094452} {\bibfield  {journal} {\bibinfo  {journal} {Phys.
  Rev. B}\ }\textbf {\bibinfo {volume} {84}},\ \bibinfo {pages} {094452}
  (\bibinfo {year} {2011})}\BibitemShut {NoStop}%
\bibitem [{\citenamefont {Parella}()}]{BrukerNMRTable}%
  \BibitemOpen
  \bibfield  {author} {\bibinfo {author} {\bibfnamefont {Teodor}\ \bibnamefont
  {Parella}},\ }\href@noop {} {\enquote {\bibinfo {title} {{eNMR}, {NMR}
  {P}eriodic {T}able},}\ }\bibinfo {note} {{BRUKER} {A}nalytik {GmbH},
  \url{http://www.bruker-nmr.de/guide/eNMR/chem/NMRnuclei.html} (accessed
  28.02.2017). According to this source, the nominal resonance frequency of the
  {$^{31}$P} nucleus in a magnetic field of 11.744 {T} is 202.404
  {MHz}.}\BibitemShut {Stop}%
\bibitem [{\citenamefont {Wang}\ \emph {et~al.}(2010)\citenamefont {Wang},
  \citenamefont {Pomjakushina}, \citenamefont {Shiroka}, \citenamefont {Deng},
  \citenamefont {Nikseresht}, \citenamefont {Rüegg}, \citenamefont {Rønnow},\
  and\ \citenamefont {Conder}}]{Wang2010}%
  \BibitemOpen
  \bibfield  {author} {\bibinfo {author} {\bibfnamefont {S.}~\bibnamefont
  {Wang}}, \bibinfo {author} {\bibfnamefont {E.}~\bibnamefont {Pomjakushina}},
  \bibinfo {author} {\bibfnamefont {T.}~\bibnamefont {Shiroka}}, \bibinfo
  {author} {\bibfnamefont {G.}~\bibnamefont {Deng}}, \bibinfo {author}
  {\bibfnamefont {N.}~\bibnamefont {Nikseresht}}, \bibinfo {author}
  {\bibfnamefont {Ch.}\ \bibnamefont {Rüegg}}, \bibinfo {author}
  {\bibfnamefont {H.~M.}\ \bibnamefont {Rønnow}}, \ and\ \bibinfo {author}
  {\bibfnamefont {K.}~\bibnamefont {Conder}},\ }\bibfield  {title} {\enquote
  {\bibinfo {title} {Crystal growth and characterization of the dilutable
  frustrated spin-ladder compound {Bi(Cu$_{1-x}$Zn$_x$)$_2$PO$_6$}},}\ }\href
  {\doibase 10.1016/j.jcrysgro.2010.09.074} {\bibfield  {journal} {\bibinfo
  {journal} {J. Cryst. Growth}\ }\textbf {\bibinfo {volume} {313}},\ \bibinfo
  {pages} {51--55} (\bibinfo {year} {2010})}\BibitemShut {NoStop}%
\bibitem [{\citenamefont {Casola}(2013)}]{DissCasola}%
  \BibitemOpen
  \bibfield  {author} {\bibinfo {author} {\bibfnamefont {F.}~\bibnamefont
  {Casola}},\ }\emph {\bibinfo {title} {Aspects of quantum magnetism in quasi
  one-dimensional materials: an {NMR} study}},\ \href {\doibase
  10.3929/ethz-a-010005724} {Ph.D. thesis},\ \bibinfo  {school} {ETH Zürich},
  \bibinfo {address} {Z\"urich, Switzerland} (\bibinfo {year}
  {2013})\BibitemShut {NoStop}%
\bibitem [{\citenamefont {Clark}\ \emph {et~al.}(1995)\citenamefont {Clark},
  \citenamefont {Hanson}, \citenamefont {Lefloch},\ and\ \citenamefont
  {S\'{e}gransan}}]{Clark1995}%
  \BibitemOpen
  \bibfield  {author} {\bibinfo {author} {\bibfnamefont {W.~G.}\ \bibnamefont
  {Clark}}, \bibinfo {author} {\bibfnamefont {M.~E.}\ \bibnamefont {Hanson}},
  \bibinfo {author} {\bibfnamefont {F.}~\bibnamefont {Lefloch}}, \ and\
  \bibinfo {author} {\bibfnamefont {P.}~\bibnamefont {S\'{e}gransan}},\
  }\bibfield  {title} {\enquote {\bibinfo {title} {Magnetic resonance spectral
  reconstruction using frequency-shifted and summed {F}ourier transform
  processing},}\ }\href {\doibase 10.1063/1.1145643} {\bibfield  {journal}
  {\bibinfo  {journal} {Rev. Sci. Instrum.}\ }\textbf {\bibinfo {volume}
  {66}},\ \bibinfo {pages} {2453--2464} (\bibinfo {year} {1995})}\BibitemShut
  {NoStop}%
\bibitem [{\citenamefont {Nijboer}\ and\ \citenamefont
  {de~Wette}(1957)}]{Nijboer1957}%
  \BibitemOpen
  \bibfield  {author} {\bibinfo {author} {\bibfnamefont {B.~R.~A.}\
  \bibnamefont {Nijboer}}\ and\ \bibinfo {author} {\bibfnamefont {F.~W.}\
  \bibnamefont {de~Wette}},\ }\bibfield  {title} {\enquote {\bibinfo {title}
  {On the calculation of lattice sums},}\ }\href {\doibase
  10.1016/S0031-8914(57)92124-9} {\bibfield  {journal} {\bibinfo  {journal}
  {Physica}\ }\textbf {\bibinfo {volume} {23}},\ \bibinfo {pages} {309--321}
  (\bibinfo {year} {1957})}\BibitemShut {NoStop}%
\bibitem [{\citenamefont {de~Wette}\ and\ \citenamefont
  {Schacher}(1965)}]{DeWette1965}%
  \BibitemOpen
  \bibfield  {author} {\bibinfo {author} {\bibfnamefont {F.~W.}\ \bibnamefont
  {de~Wette}}\ and\ \bibinfo {author} {\bibfnamefont {G.~E.}\ \bibnamefont
  {Schacher}},\ }\bibfield  {title} {\enquote {\bibinfo {title} {Internal
  {F}ield in {G}eneral {D}ipole {L}attices},}\ }\href {\doibase
  10.1103/PhysRev.137.A78} {\bibfield  {journal} {\bibinfo  {journal} {Phys.
  Rev.}\ }\textbf {\bibinfo {volume} {137}},\ \bibinfo {pages} {A78--A91}
  (\bibinfo {year} {1965})}\BibitemShut {NoStop}%
\bibitem [{\citenamefont {Albuquerque}\ \emph {et~al.}(2007)\citenamefont
  {Albuquerque}, \citenamefont {Alet}, \citenamefont {Corboz}, \citenamefont
  {Dayal}, \citenamefont {Feiguin}, \citenamefont {Fuchs}, \citenamefont
  {Gamper}, \citenamefont {Gull}, \citenamefont {G\"urtler}, \citenamefont
  {Honecker}, \citenamefont {Igarashi}, \citenamefont {K\"orner}, \citenamefont
  {Kozhevnikov}, \citenamefont {L\"auchli}, \citenamefont {Manmana},
  \citenamefont {Matsumoto}, \citenamefont {McCulloch}, \citenamefont {Michel},
  \citenamefont {Noack}, \citenamefont {Paw\l{}owski}, \citenamefont {Pollet},
  \citenamefont {Pruschke}, \citenamefont {Schollw\"ock}, \citenamefont {Todo},
  \citenamefont {Trebst}, \citenamefont {Troyer}, \citenamefont {Werner},\ and\
  \citenamefont {Wessel}}]{Albuquerque2007}%
  \BibitemOpen
  \bibfield  {author} {\bibinfo {author} {\bibfnamefont {A.~F.}\ \bibnamefont
  {Albuquerque}}, \bibinfo {author} {\bibfnamefont {F.}~\bibnamefont {Alet}},
  \bibinfo {author} {\bibfnamefont {P.}~\bibnamefont {Corboz}}, \bibinfo
  {author} {\bibfnamefont {P.}~\bibnamefont {Dayal}}, \bibinfo {author}
  {\bibfnamefont {A.}~\bibnamefont {Feiguin}}, \bibinfo {author} {\bibfnamefont
  {S.}~\bibnamefont {Fuchs}}, \bibinfo {author} {\bibfnamefont
  {L.}~\bibnamefont {Gamper}}, \bibinfo {author} {\bibfnamefont
  {E.}~\bibnamefont {Gull}}, \bibinfo {author} {\bibfnamefont {S.}~\bibnamefont
  {G\"urtler}}, \bibinfo {author} {\bibfnamefont {A.}~\bibnamefont {Honecker}},
  \bibinfo {author} {\bibfnamefont {R.}~\bibnamefont {Igarashi}}, \bibinfo
  {author} {\bibfnamefont {M.}~\bibnamefont {K\"orner}}, \bibinfo {author}
  {\bibfnamefont {A.}~\bibnamefont {Kozhevnikov}}, \bibinfo {author}
  {\bibfnamefont {A.}~\bibnamefont {L\"auchli}}, \bibinfo {author}
  {\bibfnamefont {S.~R.}\ \bibnamefont {Manmana}}, \bibinfo {author}
  {\bibfnamefont {M.}~\bibnamefont {Matsumoto}}, \bibinfo {author}
  {\bibfnamefont {I.~P.}\ \bibnamefont {McCulloch}}, \bibinfo {author}
  {\bibfnamefont {F.}~\bibnamefont {Michel}}, \bibinfo {author} {\bibfnamefont
  {R.~M.}\ \bibnamefont {Noack}}, \bibinfo {author} {\bibfnamefont
  {G.}~\bibnamefont {Paw\l{}owski}}, \bibinfo {author} {\bibfnamefont
  {L.}~\bibnamefont {Pollet}}, \bibinfo {author} {\bibfnamefont
  {T.}~\bibnamefont {Pruschke}}, \bibinfo {author} {\bibfnamefont
  {U.}~\bibnamefont {Schollw\"ock}}, \bibinfo {author} {\bibfnamefont
  {S.}~\bibnamefont {Todo}}, \bibinfo {author} {\bibfnamefont {S.}~\bibnamefont
  {Trebst}}, \bibinfo {author} {\bibfnamefont {M.}~\bibnamefont {Troyer}},
  \bibinfo {author} {\bibfnamefont {P.}~\bibnamefont {Werner}}, \ and\ \bibinfo
  {author} {\bibfnamefont {S.}~\bibnamefont {Wessel}} (\bibinfo {collaboration}
  {ALPS collaboration}),\ }\bibfield  {title} {\enquote {\bibinfo {title} {The
  {ALPS} project release 1.3: {O}pen-source software for strongly correlated
  systems},}\ }\href {\doibase 10.1016/j.jmmm.2006.10.304} {\bibfield
  {journal} {\bibinfo  {journal} {J. Magn. Magn. Mater.}\ }\textbf {\bibinfo
  {volume} {310}},\ \bibinfo {pages} {1187--1193} (\bibinfo {year}
  {2007})}\BibitemShut {NoStop}%
\bibitem [{\citenamefont {Bauer}\ \emph {et~al.}(2011)\citenamefont {Bauer},
  \citenamefont {Carr}, \citenamefont {Evertz}, \citenamefont {Feiguin},
  \citenamefont {Freire}, \citenamefont {Fuchs}, \citenamefont {Gamper},
  \citenamefont {Gukelberger}, \citenamefont {Gull}, \citenamefont {Guertler},
  \citenamefont {Hehn}, \citenamefont {Igarashi}, \citenamefont {Isakov},
  \citenamefont {Koop}, \citenamefont {Ma}, \citenamefont {Mates},
  \citenamefont {Matsuo}, \citenamefont {Parcollet}, \citenamefont
  {Paw\l{}owski}, \citenamefont {Picon}, \citenamefont {Pollet}, \citenamefont
  {Santos}, \citenamefont {Scarola}, \citenamefont {Schollw\"ock},
  \citenamefont {Silva}, \citenamefont {Surer}, \citenamefont {Todo},
  \citenamefont {Trebst}, \citenamefont {Troyer}, \citenamefont {Wall},
  \citenamefont {Werner},\ and\ \citenamefont {Wessel}}]{Bauer2011}%
  \BibitemOpen
  \bibfield  {author} {\bibinfo {author} {\bibfnamefont {B.}~\bibnamefont
  {Bauer}}, \bibinfo {author} {\bibfnamefont {L.~D.}\ \bibnamefont {Carr}},
  \bibinfo {author} {\bibfnamefont {H.~G.}\ \bibnamefont {Evertz}}, \bibinfo
  {author} {\bibfnamefont {A.}~\bibnamefont {Feiguin}}, \bibinfo {author}
  {\bibfnamefont {J.}~\bibnamefont {Freire}}, \bibinfo {author} {\bibfnamefont
  {S.}~\bibnamefont {Fuchs}}, \bibinfo {author} {\bibfnamefont
  {L.}~\bibnamefont {Gamper}}, \bibinfo {author} {\bibfnamefont
  {J.}~\bibnamefont {Gukelberger}}, \bibinfo {author} {\bibfnamefont
  {E.}~\bibnamefont {Gull}}, \bibinfo {author} {\bibfnamefont {S.}~\bibnamefont
  {Guertler}}, \bibinfo {author} {\bibfnamefont {A.}~\bibnamefont {Hehn}},
  \bibinfo {author} {\bibfnamefont {R.}~\bibnamefont {Igarashi}}, \bibinfo
  {author} {\bibfnamefont {S.~V.}\ \bibnamefont {Isakov}}, \bibinfo {author}
  {\bibfnamefont {D.}~\bibnamefont {Koop}}, \bibinfo {author} {\bibfnamefont
  {P.~N.}\ \bibnamefont {Ma}}, \bibinfo {author} {\bibfnamefont
  {P.}~\bibnamefont {Mates}}, \bibinfo {author} {\bibfnamefont
  {H.}~\bibnamefont {Matsuo}}, \bibinfo {author} {\bibfnamefont
  {O.}~\bibnamefont {Parcollet}}, \bibinfo {author} {\bibfnamefont
  {G.}~\bibnamefont {Paw\l{}owski}}, \bibinfo {author} {\bibfnamefont {J.~D.}\
  \bibnamefont {Picon}}, \bibinfo {author} {\bibfnamefont {L.}~\bibnamefont
  {Pollet}}, \bibinfo {author} {\bibfnamefont {E.}~\bibnamefont {Santos}},
  \bibinfo {author} {\bibfnamefont {V.~W.}\ \bibnamefont {Scarola}}, \bibinfo
  {author} {\bibfnamefont {U.}~\bibnamefont {Schollw\"ock}}, \bibinfo {author}
  {\bibfnamefont {C.}~\bibnamefont {Silva}}, \bibinfo {author} {\bibfnamefont
  {B.}~\bibnamefont {Surer}}, \bibinfo {author} {\bibfnamefont
  {S.}~\bibnamefont {Todo}}, \bibinfo {author} {\bibfnamefont {S.}~\bibnamefont
  {Trebst}}, \bibinfo {author} {\bibfnamefont {M.}~\bibnamefont {Troyer}},
  \bibinfo {author} {\bibfnamefont {M.~L.}\ \bibnamefont {Wall}}, \bibinfo
  {author} {\bibfnamefont {P.}~\bibnamefont {Werner}}, \ and\ \bibinfo {author}
  {\bibfnamefont {S.}~\bibnamefont {Wessel}} (\bibinfo {collaboration} {ALPS
  collaboration}),\ }\bibfield  {title} {\enquote {\bibinfo {title} {The {ALPS}
  project release 2.0: open source software for strongly correlated systems},}\
  }\href {\doibase 10.1088/1742-5468/2011/05/P05001} {\bibfield  {journal}
  {\bibinfo  {journal} {J. Stat. Mech: Theory Exp.}\ ,\ \bibinfo {pages}
  {P05001}} (\bibinfo {year} {2011})}\BibitemShut {NoStop}%
\end{thebibliography}%

\end{document}